%% file: ngc5846.tex
\documentclass{emulateapj}
\begin{document}
\input{defs}

\title{The NGC 5846 Group: Dynamics and the Luminosity Function to
$M_R=-12$}

\shortauthors{Mahdavi et al.}
\shorttitle{NGC 5846 Group}
\journalinfo{Accepted for publication in the Astronomical Journal}
\submitted{Submitted May 10,2005 ; Accepted June 27, 2005}

\author{Andisheh Mahdavi}
\affil{Institute for Astronomy, University of Hawaii, 2680 Woodlawn Drive, Honolulu, HI, USA}
\author{Neil Trentham}
\affil{Institute of Astronomy, Madingley Road, Cambridge CB3 OHA, UK}
\and
\author{R. Brent Tully}
\affil{Institute for Astronomy, University of Hawaii, 2680 Woodlawn Drive, Honolulu, HI, USA}

\newcommand{\ntotal}{324}
\newcommand{\nmemb}{199}
\newcommand{\nsdss}{64}
\newcommand{\ndEN}{61}
\newcommand{\npthree}{125}
\newcommand{\nspecmemb}{83}
\newcommand{\nvirial}{87}
\newcommand{\nnew}{19}
\newcommand{\nbrightpthree}{38}
\newcommand{\nfaintearly}{77}
\newcommand{\nfaintlate}{27}


\begin{abstract} 
We conduct a photometric and spectroscopic survey of a 10
sq. deg. region surrounding the nearby NGC 5846 group of galaxies,
using the Canada-France-Hawaii and Keck I telescopes to study the
population of dwarf galaxies as faint as $M_R=-10$.  Candidates are
identified on the basis of quantitative surface brightness and
qualitative morphological criteria.  Spectroscopic follow up and a
spatial correlation analysis provide the basis for affirming group
memberships.  Altogether, \ntotal\ candidates are identified and
\nspecmemb\ have spectroscopic membership confirmation.  We argue on
statistical grounds that a total $251 \pm 10$ galaxies in our sample
are group members.  The observations, together with archival Sloan
Digital Sky Survey, \emph{ROSAT}, \emph{XMM-Newton}, and \emph{ASCA}
data, suggest that the giant ellipticals NGC 5846 and NGC 5813 are the
dominant components of subgroups separated by 600~kpc in projection
and embedded in a 1.6~Mpc diameter dynamically evolved halo.  The
galaxy population is overwhelmingly early type.  The group velocity
dispersion is 322~km s\m, its virial mass is $8.4 \times 10^{13}
M_{\odot}$, and $M/L_R = 320 M_{\odot}/L_{\odot}$.  The ratio of
dwarfs to giants is large compared with other environments in the
Local Supercluster studied and, correspondingly, the luminosity
function is relatively steep, with a faint end Schechter function
slope of $\alpha_d = -1.3 \pm 0.1$ (statistical) $\pm 0.1$
(systematic) at our completeness limit of $M_R = -12$.
\end{abstract} 

\section{Introduction} 

The mass function of dark matter halos (i.e., their abundance as a
function of mass) is an important ingredient in constraining 
cosmological parameters using galaxies and clusters of galaxies
\nocite{Frenk90,Kauffmann93,Haiman01,Reiprich02,Hoekstra02}({Frenk} {et~al.} 1990; {Kauffmann} \& {White} 1993; {Haiman} {et~al.} 2001; {Reiprich} \& {B{\" o}hringer} 2002; {Hoekstra} {et~al.} 2002). Very
often, an assumed theoretical formula for the mass function is used to
fit observations so that cosmological quantities such as the mean
matter density of the universe $\Omega_m$ may be constrained
\nocite{Kochanek01,Haiman01,Henry04b}({Kochanek} {et~al.} 2001; {Haiman} {et~al.} 2001; {Henry} 2004). However, almost all these
methods neglect the fact that the abundance of small, $10^9-10^{10}
M_\odot$ halos is poorly understood. In fact, the very N-body
simulations used to derive forms for this fitting function,
e.g. \nocite{Sheth99}{Sheth} \& {Tormen} (1999), often produce an abundance of such halos
(sometimes referred to ``satellite galaxies'') that is far in excess
of the observations \nocite{Moore99b,Kazantzidis04,vandenBosch05}({Moore} {et~al.} 1999; {Kazantzidis} {et~al.} 2004; {van den Bosch} {et~al.} 2005). One
possibility is that the relationship between the light and mass
distributions is poorly understood at small masses \nocite{Gao04}({Gao} {et~al.} 2004)---at
the scale of dwarf galaxies. It is therefore critically important to
constrain the \emph{luminosity} function of these galaxies
\nocite{Trentham02,Tully02,Trentham05}({Trentham} \& {Tully} 2002; {Tully} {et~al.} 2002; {Trentham} {et~al.} 2005), thus providing a crucial check
on the simulations that attempt to model galaxy formation.

It is becoming clear that the relationship between the amount of
starlight (or gas) that we see and the mass of dwarf halos is not
simple.  There is evidence that the amount of light associated with
halo mass varies strongly with environment \nocite{Tully05}({Tully} 2005).  It may be
that there are dark matter halos that have retained baryons in only
undetectably small amounts.  Various astrophysical processes can lead
to a separation of dark matter and baryons, e.g., the ram pressure
stripping processes such those observed in the famous ``bullet
cluster'' 1E 0657-56 \nocite{Markevitch02}({Markevitch} {et~al.} 2002).  The relationship between
the luminosity function and the mass function might be complex, but
the luminosity function may nevertheless retain signatures of specific
astrophysical processes or of the underlying dark matter spectrum
itself.

The present article is a contribution within a long term program to
provide a better definition of the faint end of the luminosity
function of galaxies.  The general properties of the program are as
follows. First, the observations should reach very faint absolute
magnitudes.  This goal can only be achieved if the targets are nearby.
Second, good statistics are needed in conditions that provide control
of volume completion.  The issue of volume completion is addressed by
obtaining complete samples to an apparent magnitude limit in groups
selected to have minimal contamination problems.  A single group may
or may not provide adequate statistics by itself.  Observations of
numerous groups may be required.  Third, a wide variety of
environments should be sampled in order to constrain the possibility
of environmental dependencies.  This requirement imposes a need for
many nights of observations.

The program depends critically on the recent availability of panoramic
digital cameras.  The imaging material discussed in this paper was
acquired with the Canada-France-Hawaii Telescope (CFHT) 12K detector
(the predecessor of the current MegaCam).  The program also depends on
some manner of confirmation that galaxy candidates are group members,
hence relevant to the construction of the luminosity function.
Confirmation is most reliably provided by redshifts, and this paper
includes the results of spectroscopic observations of relatively
faint, low surface brightness dwarfs with the LRIS instrument on the
Keck I Telescope..

Initial attention was given to well populated clusters in both the
high density \nocite{Trentham98a,Trentham98b}({Trentham} 1998b, 1998a) and low density
\nocite{Trentham01}({Trentham} {et~al.} 2001, hereafter TTV01) regimes.  These observations
suggested that there are significant variations with environment
\nocite{Tully02}({Tully} {et~al.} 2002)---that the luminosity function of dwarf galaxies rises
steeply in dense environments, but remains flat in the field.  These
results prompted an exploratory sampling of a wide variety of
locations within the Local Supercluster with the Subaru Telescope
SuprimeCam imager \nocite{Trentham02}({Trentham} \& {Tully} 2002, hereafter TT02).  It became clear
that much more sky needed to be observed to build up meaningful
statistics.  This paper presents results from the first group in the
program to receive full coverage: a tight, well defined knot of early
type galaxies surrounding the elliptical galaxies NGC 5846 and NGC
5813.

\section{Observations}

\subsection{NGC 5846: A Well-Isolated Group of Galaxies}

This group is readily apparent because of its high density contrast
and relatively isolated location.  It has been studied as a group of
galaxies in both the X-ray and the optical
\nocite{Tully87,Haynes91,Nolthenius93,Giurcin00,Trinchieri02,Mulchaey03}({Tully} 1987; {Haynes} \& {Giovanelli} 1991; {Nolthenius} 1993; {Giuricin} {et~al.} 2000; {Trinchieri} \& {Goudfrooij} 2002; {Mulchaey} {et~al.} 2003).
The distance to the system is taken to be 26.1~Mpc from an average
over several sources, but most heavily reflective of the Surface
Brightness Fluctuation measurement of \nocite{Tonry01}{Tonry} {et~al.} (2001).  These authors
give a distance to NGC~5846 itself of $25 \pm 4$~Mpc and a distance to
NGC~5813 of $32 \pm 3$~Mpc, estimates we consider compatible with the
two being at a common distance.  The group members are overwhelmingly
of early type, dominated by the ellipticals NGC~5846 ($M_R=-22.5$) and
NGC~5813 ($M_R=-22.2$).  Within this distance, the group is the third
most massive knot of early type galaxies (after the Virgo and Fornax
clusters).  The virial mass based on velocity information to be
discussed in a later section, is $8 \times 10^{13} M_{\odot}$, and the
ratio of mass to light at R band is $320 M_{\odot}/L_{\odot}$.

\begin{figure}
\resizebox{3.5in}{!}{\includegraphics{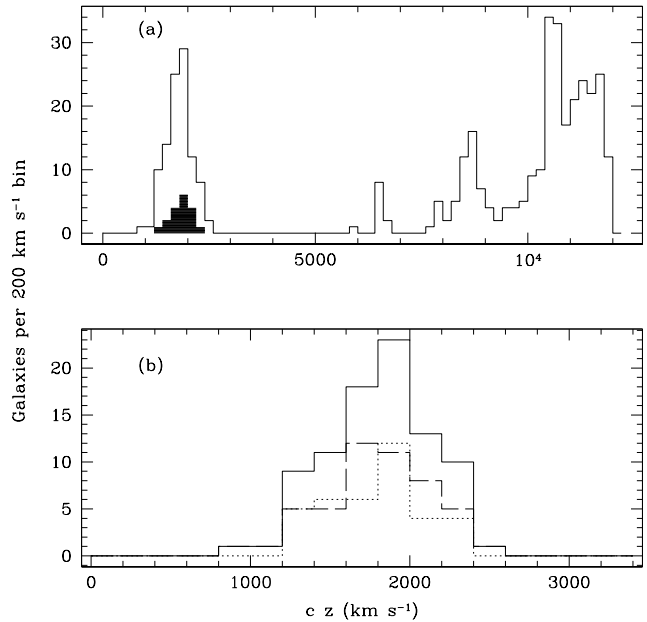}}
\figcaption{ \small Histogram of velocities of galaxies within $3^{\circ}$ of
NGC~5846.  (\emph{a}) The unfilled histogram shows SDSS and NED
galaxies; the filled histogram shows the velocities measured in our
Keck survey. (\emph{b}) All known velocities within the virial radius
(0.8 Mpc), shown as the solid histogram; the dashed histogram shows
only galaxies within 0.5 Mpc of NGC 5846, and the dotted histogram
shows only galaxies within 0.5 Mpc of NGC 5813.
\label{fig:velhist}}
\end{figure}

The isolation of the NGC~5846 Group is extremely favorable.  It lies
well off the main plane of the Local Supercluster, with no known
structure to the foreground and very little in the background until
the Hercules Supercluster at 10,000~km~s$^{-1}$.
Figure~\ref{fig:velhist}a is a histogram of all Sloan Digital Sky
Survey \nocite{Abazajian04}({Abazajian} {et~al.} 2004, hereafter SDSS) and NASA Extragalactic
Database (NED hereafter) velocities of galaxies within $3^{\circ}$
(1.4 Mpc) of the center of NGC~5846 (not including our survey below).
Remarkably, there are no galaxies at all within $3000 < c z <
6,000$~km~s$^{-1}$.

\begin{figure}
\resizebox{3.5in}{!}{\includegraphics{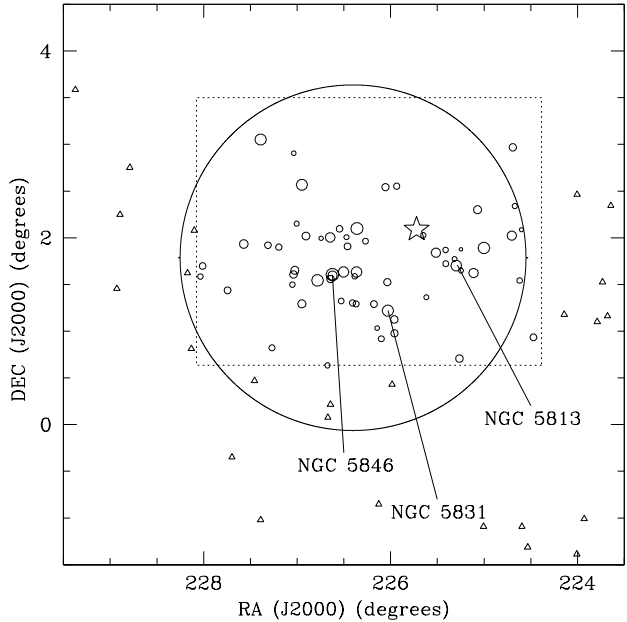}}
\figcaption{ \small Projected distribution of all galaxies with known
velocities within $900 < c z <3000~{\rm km~s}^{-1}$ and within
$3^{\circ}$ of NGC~5846.  The region of the CFH
12K imaging survey is indicated by the dotted rectangle.  The circle
of radius $r_{2t}=1.85^{\circ}=0.84$ Mpc centered at 226.40, +1.79
encloses the core of the group and shows the region corresponding to
the second turnaround radius.  Galaxies outside the survey box are
represented by triangles, while galaxies inside the survey region have
circles with size proportional to the logarithm of the $R$-band
luminosity.  The stellar symbol locates a 4th magnitude star. The NGC
numbers identify the major bright elliptical galaxies in the group.
\label{fig:presurvey}}
\end{figure}

Figure \ref{fig:presurvey} shows the distribution of all but two of
the galaxies that are established to be within 3\degr\ of the group on
the basis of a measured velocity; the two remaining objects are
described only as \ion{H}{1} detections in the literature, without
optical counterparts. The shown galaxies have velocities in the range
$900 < c z < 2700$~km~s$^{-1}$ .

\begin{figure*}
\begin{center}
\resizebox{7in}{!}{\includegraphics{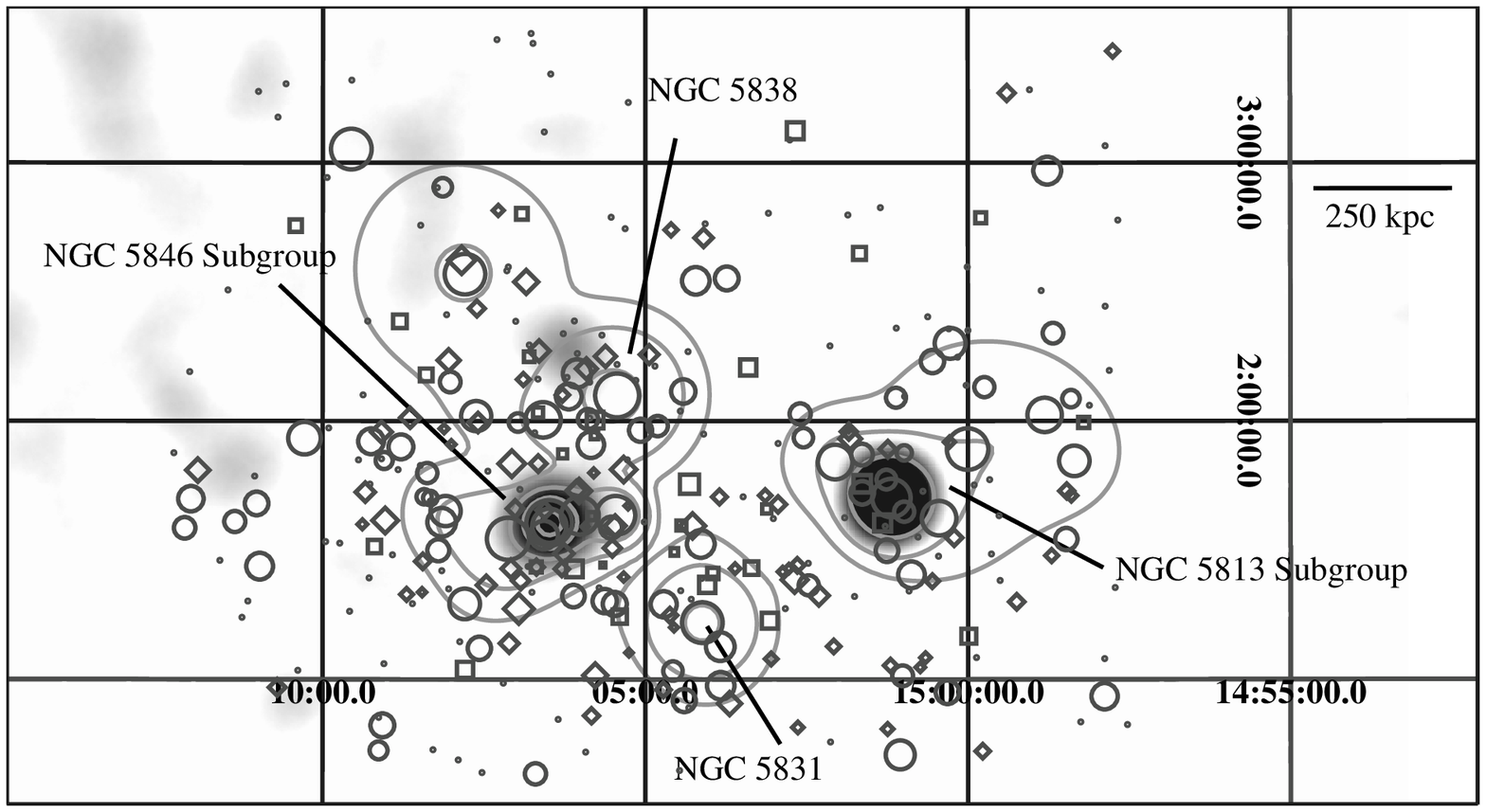}} \figcaption{ \small The NGC 5846 Group. The graylevel
image shows \emph{ROSAT} X-ray emission with significance from
3$\sigma$ (the lightest features) to 5$\sigma$ (the darkest
features). Spectroscopically confirmed, unconfirmed priority 1, and
unconfirmed priority 2 member galaxies are marked with circles,
squares, and diamonds, respectively. The size of each symbol is
proportional to the negative of the $R$ magnitude. Unconfirmed priority
3 galaxies are marked with dots of the same size. The contours show
the adaptively smoothed $R$-band light distribution of the members, in
units of 0.1, 0.5, 1, 5, and 10 $L^R_\sun$ pc$^{-2}$. The Gaussian
smoothing length for each galaxy is the distance to the third-nearest
member. \label{fig:bigmap}}
\end{center}
\end{figure*}

Over the range of environments that will be explored in this program,
the NGC~5846 Group lies in the regime of high density and intermediate
mass.  It possesses 4 galaxies brighter than $L^{\star}$ (3 E/S0 and
an Sb).  In this paper, \ntotal\ galaxies are identified as probable or
possible members, extending in faintness down to $M_R \sim -10$.

\subsection{Wide Field Imaging}

Observations of the NGC 5846 Group were made with the CFH12K CCD
camera in queue mode during 11 nights between 16 March 2002 and 10
June 2002.  An overall rectangular area of $220^{\prime} \times
180^{\prime}$ was surveyed with a small hole to avoid the glare of a
4th magnitude star.  Two control fields were observed $9^{\circ}$
north, in a region with no known objects foreground of $\sim
10,000$~km~s$^{-1}$.  The CFHT12K detector is a mosaic of 12 CCD
detectors providing a field of $42^{\prime} \times 28^{\prime}$,
oriented in this experiment with the long axis E-W.  The observations
were tiled with half-field overlaps and dithers so that gaps between
CCD chips were almost entirely covered in subsequent exposures and
most of the area was observed twice.  In total, 67 x 9 minute
exposures were taken, all in the Cousins R band, covering 10.05 square
degrees.  Two of the 11 nights were non-photometric but photometry
could be propagated across the entire survey region through the
half-field overlaps.  Seeing was 0.7--1.0 arcsec as mandated by the
queue request.  Images of the individual dwarfs will be available
online via the CFHT image cutout service (planned for the future).

Members of the NGC~5846 Group range from spectroscopically accessible
high surface brightness objects ($\mu_R < 20$ mag arcsec$^{-2}$) to
spectroscopically challenging very low surface brightness objects
($\mu_R > 20$ mag arcsec$^{-2}$).  Experience has shown that {\it
most} low luminosity galaxies are low surface brightness, although
there are exceptions \nocite{Drinkwater03}({Drinkwater} {et~al.} 2003).  Candidates can be isolated
on the basis of this property.  Morphological criteria can then be
applied to further evaluate the probability of group membership.  The
details of the procedures to chose candidates have been described in
TTV01 and TT02.  The essence of the process is a culling of the very
large background population with a concentration index threshold.
Known dwarfs lie in a distinctly lower concentration index regime that
giant galaxies.  Nonetheless, substantial numbers of background
contaminants manage to pass the concentration index screen and must be
culled based on morphological criteria.  This latter step requires
inspection of images with the following considerations in mind.  If
there is evidence for a bulge or major bar, or spiral structure, or
tidal disruption then such objects that pass the concentration filter
are probably background.  If instead an object is diffuse except
possibly for a semi-stellar nucleation or patchy structure then it is
probably nearby, hence a group member.

In TTV01 and TT02 there is a description of a rating scheme developed
to characterize the probability of group membership for galaxies which
have passed the surface brightness criterion 
(1: probable member; 2: possible member; 3: conceivable
member; 4: almost certainly background).  
Additionally, a rating 0 is given to galaxies that were identified as
group members on the basis of redshifts before the survey began, whatever
their surface brightness, and a rating 5 is given to galaxies initially
rejected as candidates because they lie above the concentration index cut
but subsequently associated with the group by a redshift.
The availability of velocity
information for a significant fraction of the sample now permits a
reevaluation of the validity of the morphology-based rating scheme.
In \S\ref{sec:membership}\ below we show that galaxies rated 1 and 2
are essentially always found to be group members, of order half the
galaxies rated 3 are found to be group members, but almost none of the
galaxies rated 4 are group members.  Five high surface brightness galaxies 
(identified with a rating 5) are found to be members.

Before considering the SDSS and out Keck data, 318 galaxies are
identified in the survey region with ratings 0-3. After including the
SDSS, the number increases to \ntotal, because 6 priority 4-5 galaxies
not originally included in our sample are confirmed by the SDSS as
members.  These objects are identified in Table~\ref{tbl:data}.  The
magnitudes presented in this table are isophotal $R$-band magnitudes
extracted to an isophote of 25.2 mag arcsec\m.  Their distribution on
the sky is shown in Figure~\ref{fig:bigmap}.  There is a strong
enhancement in the surface number density of both confirmed members
and candidates surrounding the elliptical NGC~5846 and a secondary
enhancement surrounding the elliptical NGC~5813.

The two control fields $9^{\circ}$ N were chosen to lie off the
filament containing the NGC~5846 Group and in the direction of the
Local Void \nocite{TullyFisher87}({Tully} \& {Fisher} 1987).  No galaxies were found in these
fields that could be rated 1--3.  On the basis of the detection rate
in the NGC~5846 area, one would anticipate 9 such candidates in these
fields.

\subsection{Spectrocopic Observations}
\label{sec:velocities}

\begin{figure*}
\begin{tabular}{cc}
\resizebox{4.in}{!}{\includegraphics*[1.4in,3.5in][6.9in,7.3in]{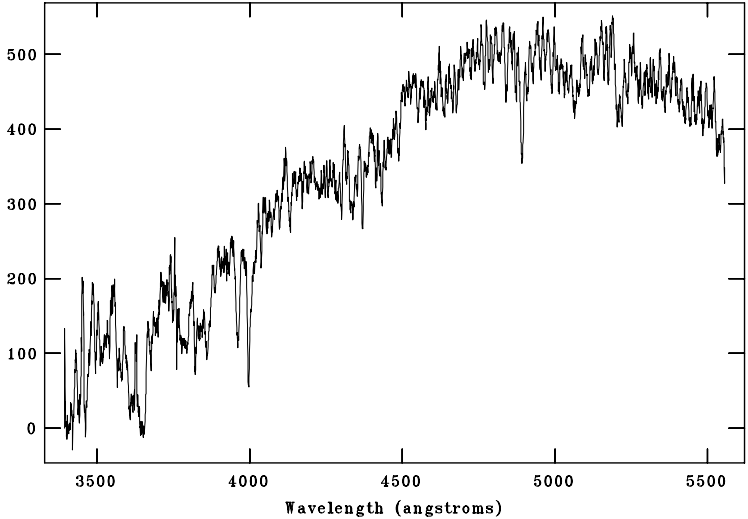}} &
\resizebox{3.in}{!}{\includegraphics{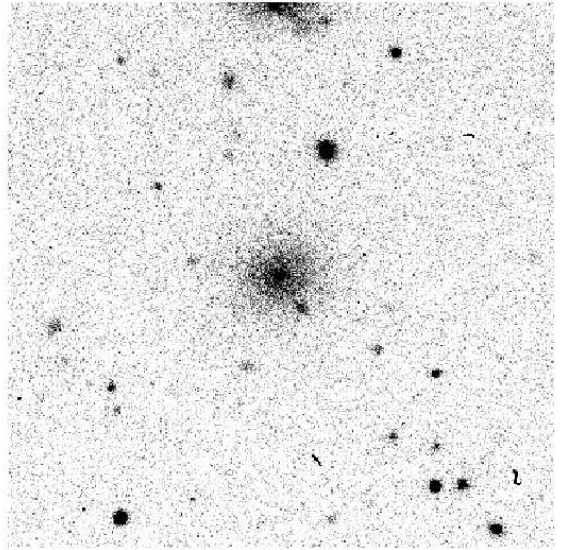}} \\
\end{tabular}
\figcaption{ \small  (\emph{left}) Blue LRIS spectrum for the member galaxy 
N5846--264, the faintest galaxy observed with the Keck I telescope
($m_R = 18.8$, $M_R=-13.4$). The spectrum quality is typical for
the Keck I sample. Visible absorption lines are redshifted \ion{Ca}{2}
lines at 3955 and 3998 \AA, H$\beta$ at 4893 \AA, and Mg at 5208
\AA. Slight H$\delta$ is visible in emission at 4111
\AA. (\emph{right}) $1.4\arcmin \times 1.4 \arcmin$ CFHT 12K image of
N5846-264.
\label{fig:samplespec}}
\end{figure*}

The NGC~5846 Group lies within the area with published spectroscopic
information from Data Release 3 of the SDSS.  The NASA/IPAC
Extragalactic Database (NED) and the SDSS provide velocities for
\nsdss\ of our targets, all of them brighter than $R \sim 17$ or $M_R
\sim -15$.  There is essentially no prior literature information for
candidates at fainter magnitudes or for those with very low surface
brightnesses.  In order to probe these regimes, we undertook
observations using the blue side of the Low Resolution Imaging
Spectrograph \nocite{Oke95}({Oke} {et~al.} 1995, LRIS hereafter), on Keck I Telescope. The
spectrograph, equipped with a 1\arcsec\ slit and a 600 lines in$^{-1}$
grating, has high blue quantum efficiency, with an overall system
throughput of 56\% at 5000 \AA.  Data were acquired on May 4, 2003 and
June 13, 2004.  Both nights were hampered by cirrus with extinction of
0.5-2 mag; nevertheless we obtained suitable spectra for 17 and 13
galaxies on the two nights, respectively, of which a total of \nnew\
are newly reported members.  The poor conditions limited attempts to
observe the faintest dwarfs; however, good spectra were acquired for
objects as faint as $R = 18.8$ or $M_R = -13.4$. An LRIS
spectrum and CFH 12K image are shown in Figure~\ref{fig:samplespec}
of the faintest galaxy with a redshift.

\subsection{Spectroscopic Membership Confirmation}
\label{sec:membership}

The availability of a large number of new velocities provides a way of
evaluating the membership rating scheme based on a concentration
parameter and a qualitative judgment based on morphology.  The
NGC~5846 Group provides an environment that is particularly
well-suited to this evaluation, because there is negligible confusion
from the foreground or near background.  Table~\ref{tbl:members}
provides a summary of how things have turned out.  We found all 26
surveyed galaxies with new velocities rated 1 (probable) and 2
(possible) to be group members.  Of the galaxies rated 4 (likely
background) only one relatively large galaxy has been revealed by
spectroscopy to be a member.  Among the thousands of galaxies in the
survey region excluded by the concentration criteria, 304 have
measured redshifts; of these, only 5 galaxies have been demonstrated
to be group members. 

Thus candidates rated 1 and 2 ought to be group members and, by
contrast, very few group members emerge among galaxies rated 4 or
those excluded because of high concentration.  That leaves the
galaxies rated 3 (conceivable member) to be considered.  We find that
velocity measurements confirmed 16 galaxies with rating 3 to be
members and 7 galaxies to be background.

In summary, the combination of the quantitative concentration
parameter and the qualitative morphological evaluation leads to good
membership discrimination.  There remains a grey area with the
candidates rated~3.  High surface brightness objects elude discovery
in the imaging survey but can be found with a spectroscopic survey.
Having said all this, the spectroscopic confirmation is complete only
to $M_R \sim -15$ and sampled only to $M_R \sim -13.3$.  There cannot
be complete confidence that the rating scheme works among the fainter
galaxies that extend down to $M_R \sim -10.5$.  It would be
surprising, though, if there is a population of high surface
brightness galaxies in the group at these faint magnitudes that makes
a significant numeric contribution to the luminosity function.

Henceforth we refer to all member galaxies with redshifts as
``spectroscopically confirmed members''; when we refer to priority 0-3
galaxies, we mean only those without a redshift.

\subsection{Indicative Membership: Spatial Correlation}
\label{sec:membership2}

\begin{figure*}
\begin{center}
\resizebox{7in}{!}{\includegraphics{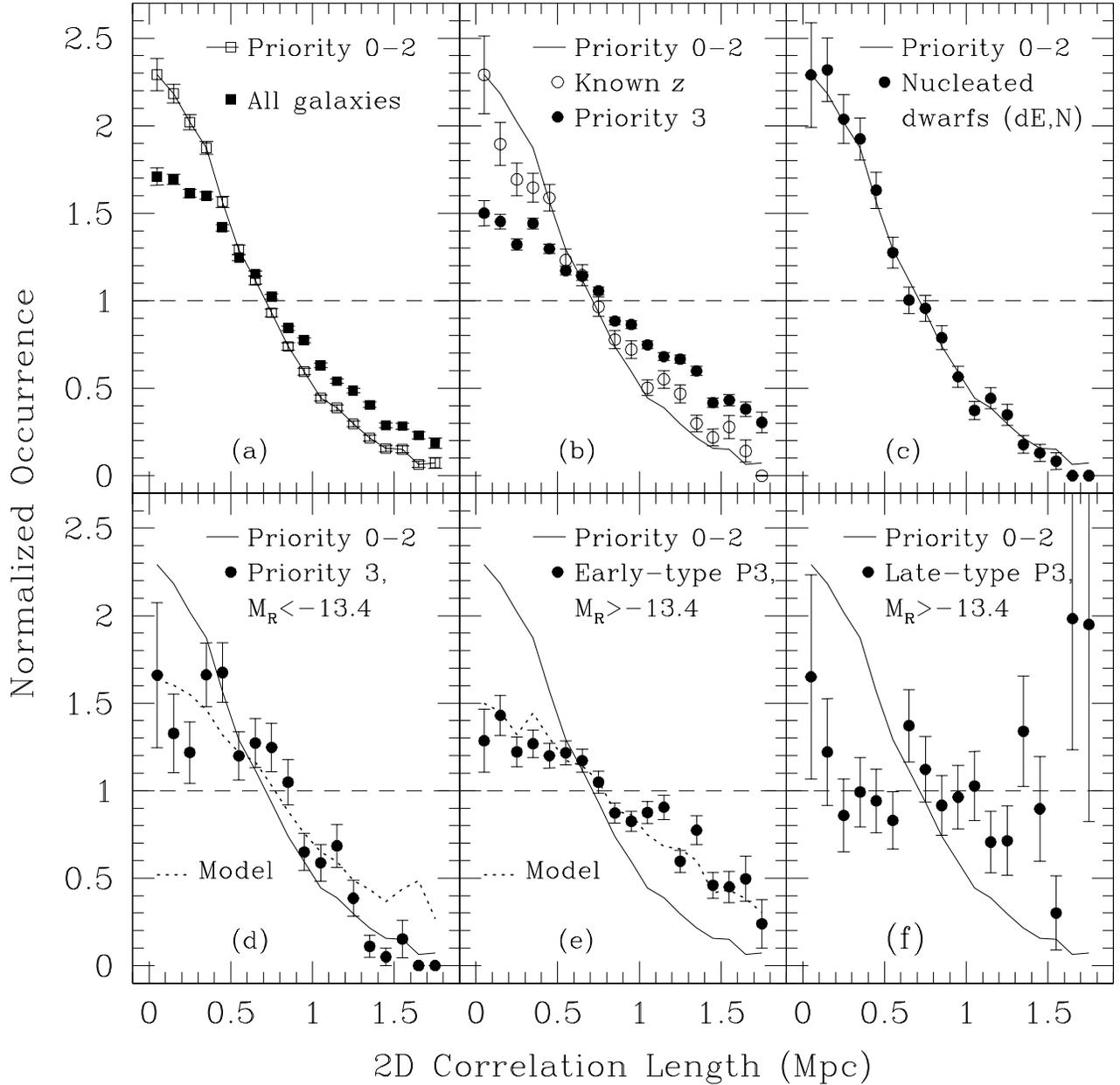}}
\end{center}
\figcaption{ \small  Two-point correlations. ({\it a}) The filled
symbols show the normalized 2-point correlation between all \ntotal\
candidates for membership in the NGC~5846 group.  The open symbols
give the correlation for the \nmemb\ galaxies with either redshift
confirmation (rating 0) or without known redshift but with membership
probability ratings 1 or 2.  ({\it b}) The open symbols give the
correlation for the \nspecmemb\ galaxies in the area of the CFHT
12K survey confirmed as members on the basis of redshifts.  The
filled symbols give the correlation for the \npthree\ galaxies with
membership probability rating 3.  ({\it c}) Filled symbols show the
correlation for \ndEN\ nucleated dwarf ellipticals.  ({\it d}) The
correlation is shown for the \nbrightpthree\ galaxies with membership
rating 3 that are brighter than $M_R=-13.4$.  The dotted curve
illustrates the correlation with a 70\% contribution consistent with
the reference correlation (the solid curve) and a 30\% random
component.  ({\it e}) The correlation achieved for \nfaintearly\
galaxies with membership rating 3 that are fainter than $M_R=-13.4$
and early morphological types.  The dotted curve presents the
correlation with a 50\% contribution consistent with the reference
correlation (solid curve) and a 50\% random component.  ({\it f}) The
result of the correlation analysis with \nfaintlate\ galaxies with
membership rating 3, fainter than $M_R=-13.4$, and late morphological
type.  These objects show no positive correlation with the group.
\label{fig:corr}}
\end{figure*}

Concentrations of galaxies toward NGC~5846 and NGC~5813 are seen in
the galaxy projections of Fig.~\ref{fig:presurvey} and in
Fig.~\ref{fig:bigmap} introduced in the next section.  A comparison of
angular 2-point correlations among various sub-samples gives hints of
different degrees of clustering.  Sub-samples with relatively weak
correlations are probably contaminated by non-members.  The
correlation that exists among all \ntotal\ galaxies rated 0-3 is shown
in the upper left panel of Figure~\ref{fig:corr}.  The normalization
in the correlation is achieved by comparison with 1000 Monte Carlo
random populations of the area of the photometric survey.  The entire
area is overdense with respect to a fair sample of the universe so the
amplitude of the normalization is given no meaning.

Next consider the correlation shown in the same panel for \nmemb\
galaxies either with membership confirmed by redshift or rated 1--2.
The spectroscopic evidence suggests that most of the sample receiving
these rating are members.  Hence the correlation function should be
fairly representative of the true global function for the group.  We
refer to this function as the ``reference correlation'' hereafter.
The reference correlation is more peaked than the function describing
the ensemble of the candidates.  The ensemble sample must be
contaminated by objects drawn from the background.

Figure~\ref{fig:corr}b shows the correlation for the \nspecmemb\
galaxies established to be group members on the basis of velocities.
This correlation function is noisier because of smaller numbers but is
comparable to the function for the \nmemb\ member priority 0-2 sample.
The reference correlation is actually slightly steeper which tends to
confirm that the galaxies rated 1 and 2 are overwhelmingly members.  A
different situation is found in the correlation shown for the
\npthree\ galaxies rated 3.  The much flatter distribution is evidence
that quite a few galaxies rated 3 are non-members.  The spectroscopic
information already discussed revealed that only 16 of 23 candidates
rated 3 (70\%) are members.

In the Figure \ref{fig:corr}c, the correlation function is shown for
the \ndEN\ galaxies with the morphological designation dE,N, nucleated
dwarf ellipticals.  Only 12 of these have membership confirmed by
velocities.  Most lie at faint magnitudes.  The correlation analysis
suggests that, overwhelmingly, galaxies of this morphology are
members.

The spectroscopic information extends to galaxies as faint as
$M_R=-13.4$.  In total there are \nbrightpthree\ priority 3 candidates
brighter than this limit.  The correlation analysis for these galaxies
appears in Figure \ref{fig:corr}d.  The correlation is intermediate
between the reference function and that shown by the entire rating 3
sample in panel (b).  We create a simple model in which 
30\% of the galaxies in the reference correlation are replaced by
randomly distributed objects. This concoction of 70\% correlated and
30\% uncorrelated components provides a good description of the
distribution seen in the bright priority 3 galaxies. This result
agrees with the spectroscopic information available for half the
objects in question.

There is no velocity information concerning the fainter candidates,
and from the poorer correlation seen in Figures
\ref{fig:corr}e-\ref{fig:corr}f it can be inferred that many of the
fainter rating 3 targets are drawn from the background.  In fact, the
early and late morphological types rated 3 have significantly
different correlation characteristics.  The distinct distributions are
revealed in Figure \ref{fig:corr}e (early types) and \ref{fig:corr}f
(late types).  The \nfaintearly\ galaxies typed dE, dE,N, and dE/I
show the correlation seen in panel (e) that is well described by a mix
of objects that are 50\% correlated and 50\% uncorrelated (uncertainty
$\sim 10\%$).  By contrast, the late types (mostly dI, a few VLSB)
show no correlation to the group.  To summarize, the correlation
distribution of the rating 3 candidates seen in the filled symbols in
Fig.~\ref{fig:corr}b can be decomposed into 70\% group members among
those brighter than $M_R=-13.4$, 50\% group members among fainter
early types, and essentially no group members among fainter late
types.

One final consideration is morphological segregation of the galaxies.
It is well known that in groups and clusters, early-type galaxies are
more densely clustered than late-type galaxies
\nocite{Dressler80,Postman84,Helsdon03}({Dressler} 1980; {Postman} \& {Geller} 1984; {Helsdon} \& {Ponman} 2003). If the fraction of early-type
galaxies in the three priority groups differs greatly, the
morphology-density effect would significantly influence the
correlation functions. This would lead us to misinterpret differences
among the correlation functions as differences in the membership
probabilities of the galaxies. Fortunately, Table \ref{tbl:members}
shows that the morphological content of the three priority classes is
quite similar. The weighted average early type fraction in the
priority 0-2 galaxies is 0.82, and for the priority 3 galaxies it is
0.79. The difference in the correlation functions is too great to be
explained just by this small difference in the early-type fraction.

Based on this simple model, we can calculate an estimate of the total
group membership within our survey limits: 83 spectroscopically
confirmed members; 32 priority 1 members; 84 priority 2 members;
60-80\% of 14 bright priority 3 members; and 40-60\% of the 84 faint
early-type priority 3 members. The total estimate is $251 \pm 10$
group members.

\section{Structure of the NGC 5846 System}

\subsection{Broad X-ray and Optical Picture}

\begin{figure*}
\begin{tabular}{cc}
\resizebox{3.5in}{!}{\includegraphics{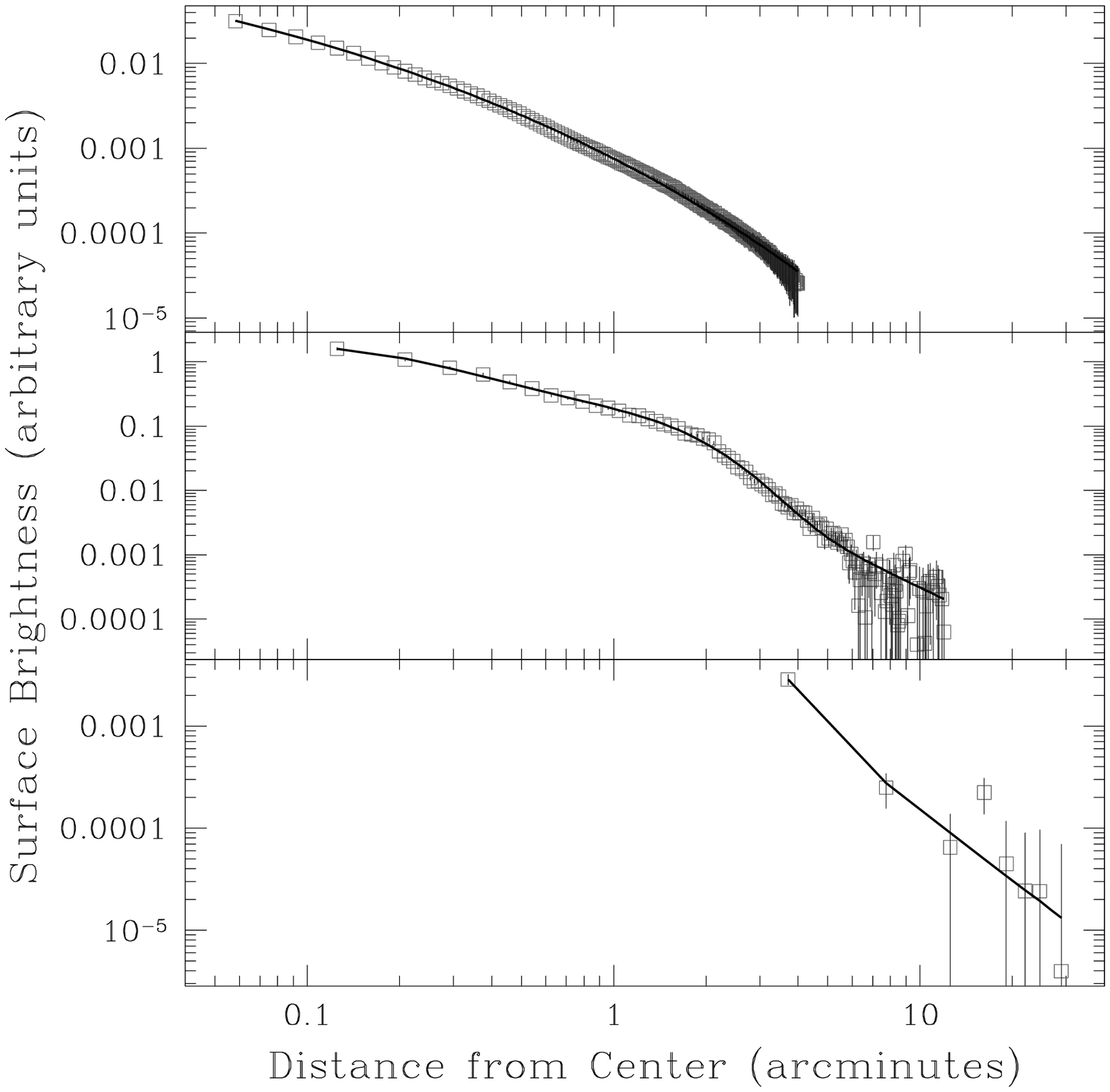}} &
\resizebox{3.5in}{!}{\includegraphics{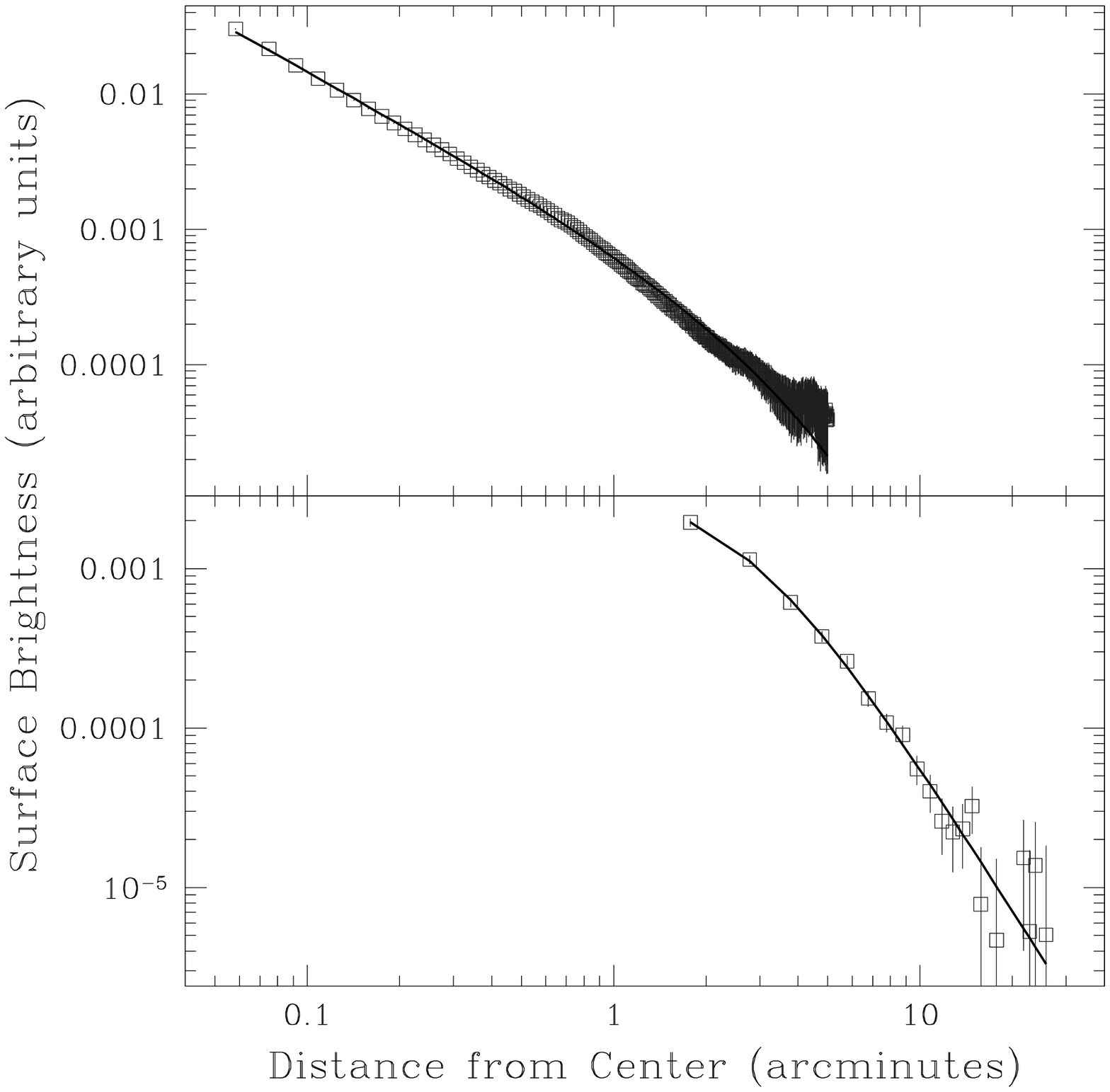}}
\end{tabular}
\figcaption{ \small (\emph{left}) From top to bottom, SDSS $g$-band,
\emph{XMM-Newton}, and RASS X-ray light distributions for NGC 5846,
along with the best-fit S\'ersic and $\beta$-models. The bottom two
figures show simultaneous fits to the X-ray data. (\emph{right}) SDSS
$g$-band and \emph{ASCA} X-ray light distribution for NGC 5813.
\label{fig:profiles}}
\end{figure*}

Combining the SDSS data with our observations, we are able to provide
a broad picture of the environment of the NGC 5846 system. As evident
in Figures \ref{fig:velhist} and \ref{fig:bigmap}, the system
possesses complex substructure, but remains remarkably well isolated
from other groups or clusters.

Another clue to the physical state of the system is the distribution
of X-ray emission across the group. In elliptical-dominated groups, a
hot, $\sim 1$ keV optically thin plasma often constitutes the largest
baryonic component of the group's mass.  Archival ROSAT, ASCA,
Chandra, and XMM-Newton observations of the NGC 5846 region exist. The
ROSAT All-Sky Survey (RASS) covers the entire 10 sq. deg. field of
our observations. However, with an average exposure time of only 355s,
only the X-ray emission closest to the two brightest galaxies, NGC
5846 and NGC 5813, is detected in the survey. We show the 0.5-2.0 keV
RASS emission in Figure \ref{fig:bigmap}. A weak X-ray source $\approx
40\arcmin$ to the north of NGC 5846 is visible, but it is not associated
with any group galaxies, and no optical sources in either the POSS or
the SDSS appear to coincide with it. Also shown is the adaptively
smoothed light distribution of the region.  Each member galaxy is
represented by a Gaussian on the sky with standard deviation equal to
the distance to the third nearest member galaxy. The Gaussians are
then added together. Figure \ref{fig:bigmap} makes it clear that the
two highest surface brightness features in the group, in both the
X-ray and the optical, are the regions within $\approx 300$ kpc of the
two brightest galaxies.

The broad picture revealed in both the optical and X-ray surface
brightness maps is unusual.  The NGC 5846 and 5813
galaxies are two roughly equal peaks of emission at both
wavelengths, so the system as a whole appears binary.
However, there are also other ellipticals nearly as bright,
such as NGC 5838 and NGC 5831, without significant X-ray emission but
with significant concentrations of smaller galaxies surrounding them.
We will explore the dynamics of the system in greater detail in
\S\ref{sec:dynamics}.

\subsection{Two Subgroups: NGC 5846 and NGC 5813}

A central question in understanding the NGC 5846 system regards the
existence of ``group-scale'' X-ray emission. Both NGC 5846 and NGC
5813 possess diffuse X-ray emission from a hot, optically thin plasma,
suggesting that they mark the center of the two largest dark matter
concentrations in the system. But is the X-ray emitting gas confined
to the galaxies only, emitting X-rays properly only as the hot
interstellar medium of the elliptical galaxies \nocite{Eskridge95a}({Eskridge} {et~al.} 1995), or
is there evidence of a true intracluster medium (ICM) in which the
galaxies are embedded? An extended ICM would make it clear that the
NGC 5846 and 5813 galaxies are not only bright, X-ray emitting
ellipticals; they are also the centers of much more massive dark
matter halos. Previous analyses have considered \emph{ROSAT} data
\nocite{Boehringer00,Ikebe02,Osmond04}({B{\" o}hringer} {et~al.} 2000; {Ikebe} {et~al.} 2002; {Osmond} \& {Ponman} 2004), but no comparisons with
high-quality surface brightness profiles (such as from the SDSS) or
discussions of the \emph{XMM-Newton} observations of NGC 5846 exist.

To evaluate the extent of the X-ray emission, we examine
\emph{XMM-Newton} data for NGC 5846 (superior in sensitivity and
field-of-view to the \emph{Chandra} data) and the \emph{ASCA} data for
NGC 5813 (where neither \emph{XMM-Newton} nor \emph{Chandra} data is
available). We extract X-ray surface brightness profiles from the
available data. In the case of \emph{ROSAT} and \emph{ASCA}, we
extract pre-calibrated photon images and exposure maps from the
\emph{ROSAT} All-Sky Survey Data
Browser\footnote{\url{http://wave.xray.mpe.mpg.de/rosat/data-browser}}
and the NASA HEASARC data
archive\footnote{\url{http://heasarc.gsfc.nasa.gov/W3Browse/}},
respectively. For \emph{XMM-Newton}, only images from the
highest-sensitivity camera on board the telescope, the EPIC pn, were
available. We use the \emph{XMM-Newton} Software Analysis System
(SAS)\footnote{\url{http://xmm.vilspa.esa.es/sas/}} to reduce these
data.

In the case of NGC 5846, the RASS contributes a useful signal outside
the $20\arcmin$ diameter field-of-view of \emph{XMM-Newton}. In the
case of NGC 5813, \emph{ASCA} provides superior data to the RASS. For
the RASS, we use the 0.5-2.0 keV data to minimize the noise from the
unrelated X-ray background. For the \emph{XMM-Newton} EPIC pn we also
use photons with 0.5-2.0 keV energies. For the NGC 5813 observations
by the \emph{ASCA}, we use the full-band images covering photon
energies 0.7-10 keV. We also extract optical surface brightness
profiles from calibrated $g$-band imaging data publicly available from
the SDSS.

The profiles appear in Figure \ref{fig:profiles}. In order that our
analysis is not affected by the point spread function (PSF) of each
instrument, we ignore data within a radius 4 times the full width of
the PSF at half-maximum; the PSF deformation of the profiles should be
negligible beyond these radii \nocite{Mohr99}({Mohr} {et~al.} 1999). We fit the profiles with
simple azimuthally symmetric models to evaluate their extent. While
the models may not be correct in detail, they are useful tools in
characterizing the distribution of light in the group. To fit the
SDSS $g$-band optical light distribution, we use a modified S\'ersic
\nocite{Sersic68,Trujillo04}({S\'ersic} 1968; {Trujillo} {et~al.} 2004) profile:
\begin{equation}
\Sigma_g(r) = \Sigma_0 (r/r_e)^\gamma 
\exp{\left[ -b \left(r/r_e\right)^\alpha \right]},
\end{equation}
where $r$ is the projected distance to the galaxy center, $r_e$ is the
half-light radius, b is a constant, and $\alpha$ and $\gamma$ are the
characteristic slopes. The case $\gamma=0,\alpha=1/4$ corresponds to
the widely known \nocite{deVaucouleurs76}{de Vaucouleurs} {et~al.} (1976) profile. The S\'ersic profile
($\gamma=0$, $\alpha$ free) is known to fit a subset of the bright
elliptical galaxies observed with the Hubble Space Telescope (HST).
Others, however, require a nonzero inner slope $\gamma$ for an
acceptable fit \nocite{Trujillo04}({Trujillo} {et~al.} 2004)
\footnote{In \protect\nocite{Trujillo04}{Trujillo} {et~al.} (2004), these so-called
``core-S\'ersic'' galaxies have inner profiles of the form
$[1+(r/r_b)^\delta]^{\gamma/\delta}$. However, typical values of $r_b$
are $50-100$ pc, requiring much finer resolution to resolve than the
data we discuss here possess. Therefore we approximate the inner region
of the \nocite{Trujillo04}{Trujillo} {et~al.} (2004) profile as a simple power law with slope
$\gamma$.}.

The X-ray light distribution is characterized by a double
$\beta$-model \nocite{Mohr99}({Mohr} {et~al.} 1999):
\begin{eqnarray}
\Sigma_x(r) & = & \Sigma_1 (1+r^2/r_1^2)^{-3 \beta_1 + 1/2} \nonumber \\ 
& +&  
\Sigma_2 (1 + r^2/r_2^2)^{-3 \beta_2 + 1/2}.
\end{eqnarray}
The single $\beta$-model was originally used to describe data observed
by the \emph{Einstein} observatory \nocite{Jones84}({Jones} \& {Forman} 1984). Higher resolution
\emph{Chandra} observations have since revealed that emission from
relaxed clusters is often more complex and requires either a broken
power law or the two-component $\beta$-model shown above
\nocite{Buote04}({Buote} \& {Lewis} 2004).

The fit results in Table \ref{tbl:fits} suggest substantial
differences between the two galaxies. The $\gamma=0$ optical fit for
NGC 5846 is of good quality. The brightest group member is consistent
with being a \nocite{deVaucouleurs76}{de Vaucouleurs} {et~al.} (1976) model galaxy ($\alpha = 1/4$). For
NGC 5813 the pure $\gamma=0$ S\'ersic profile is not a good fit. With
$\chi^2/\nu = 364/291$, the fit is rejected at better than 99.7\%
confidence. However, with $\gamma$ free, we obtain a good fit. Thus,
the dominant galaxies in the group, though possessing similar Hubble
types (E0 and E1 for NGC 5846 and 5813, respectively), exhibit
different light profiles in detail. NGC 5813's inner stellar surface
brightness profile is consistent with a power law of index $1.2$, but
that of NGC 5846 flattens near the center. Such variations in profile
shapes are also present in field ellipticals observed by HST
\nocite{Trujillo04}({Trujillo} {et~al.} 2004), though the reasons for the existence of two
populations are not yet clear.

One clue to understanding the differences in the light profiles of the
two galaxies may lie in their merging histories. Detailed HST
observations of the central few arcsec in NGC 5813 reveal a dusty
circumnuclear disk \nocite{Tran01}({Tran} {et~al.} 2001). The inner region of NGC 5846
contains no disk, but X-ray and optical filamentary structures instead
\nocite{Goudfrooij98,Tran01,Trinchieri02}({Goudfrooij} \& {Trinchieri} 1998; {Tran} {et~al.} 2001; {Trinchieri} \& {Goudfrooij} 2002). One interpretation is that
the filaments in NGC 5846 are relics of interaction with smaller
galaxies \nocite{Goudfrooij98}({Goudfrooij} \& {Trinchieri} 1998). The fact that NGC 5813 contains an
undisturbed circumnuclear disk suggests that it may have had fewer
recent mergers than NGC 5846. For this reason NGC 5813 retains a
steeper, more undisturbed stellar profile than does its sister galaxy.

The X-ray data reveal that the intracluster medium is more extended
than the optical light distribution. We obtain acceptable fits for a
double $\beta$-model in NGC 5846, while for NGC 5813 a single
$\beta$-model suffices. The ratio of the X-ray to the optical
half-light radii for NGC 5846 and NGC 5813 are $1.3 \pm 0.3$ and $3.1
\pm 0.8$, respectively; the inferred 90\% light radii have ratios of
$>2.2$ and $>8.2$, respectively. Interestingly, though NGC 5813 is
optically less bright NGC 5846, it contains a more luminous and
extended X-ray halo.

\begin{figure}
\resizebox{3.5in}{!}{\includegraphics{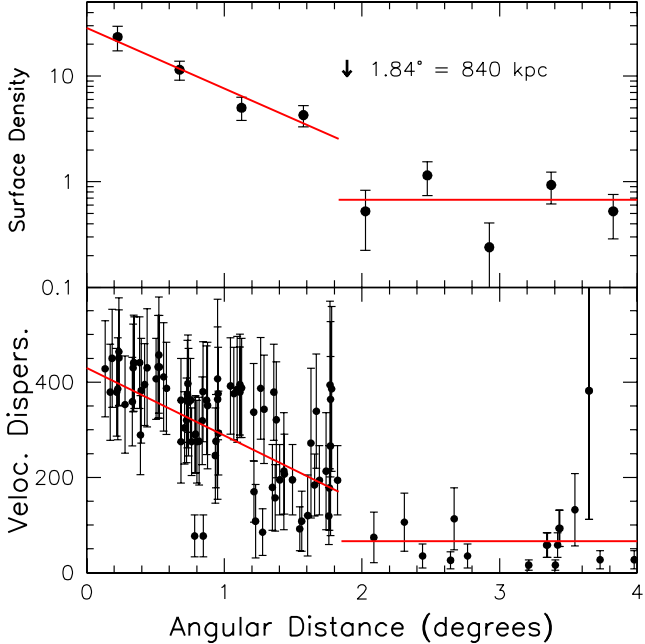}} \figcaption{ \small Radial gradients.  {\it Top:}
surface number density dependence on radius for galaxies with observed
velocities and $R<17.2$.  The surface number density drops smoothly
with radius out to $\sim 1.8^{\circ}$ then drops abruptly to a roughly
constant value. {\it Bottom:} Velocity dispersion as a function of
radius. Each data point represents the velocity dispersion of all
galaxies within 0.5 Mpc of a single galaxy; thus, the data points are
not independent. The sloping line within 1.8\degr\ is fit to velocity
dispersions averaged over the 4 annular bins of the top panel. The
flat line, showing the mean velocity dispersion outside 1.8\degr, lies
at 66 km s\m.
\label{fig:radgrad}}
\end{figure}

\begin{figure*}
\resizebox{7in}{!}{\includegraphics{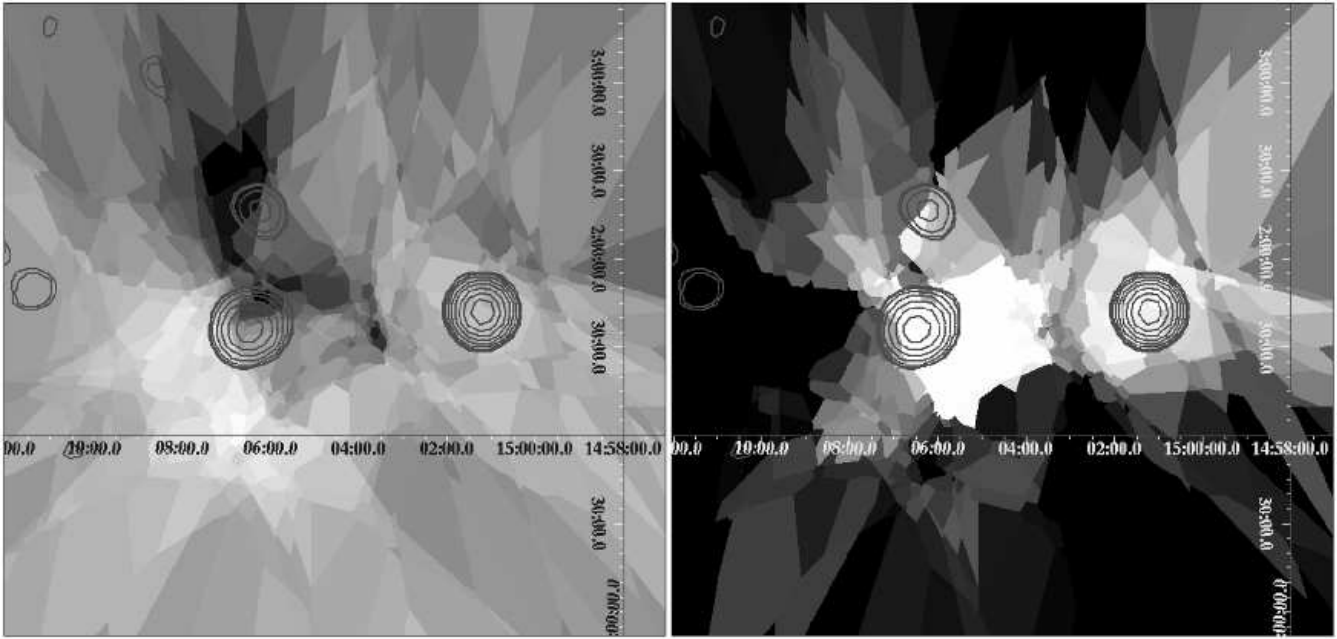}}
\figcaption{ \small Velocity and velocity dispersion maps of the NGC 5846
system. In both maps, contours indicate X-ray emission at 2, 3, 4, and
5,  6, and 7 $\sigma$ significance. (\emph{Left}) The grayscale map shows the
mean galaxy velocity of the 10 nearest members; black corresponds to
1560 km s\m, and white corresponds to 2130 km s\m. (\emph{Right}) The
grayscale map shows the velocity dispersion of the 10 nearest members;
black corresponds to 132 km s\m, and white corresponds to 562 km
s\m. \label{fig:meansigmap}}
\end{figure*}

\subsection{Dynamics}
\label{sec:dynamics}

On dynamical grounds, the NGC~5846 Group can be expected to consist of
an evolved core surrounded by an infall region.  Given the mass of the
group, the infall region is expected to extend to $\sim 10^{\circ}$
radius, considerably beyond the boundaries of the survey region
illustrated in Fig.~\ref{fig:presurvey}.  The interest of the current
study is with the dense, dynamically evolved core.  The dimensions of
that core can be inferred from the radial density and velocity
distribution of galaxies in the vicinity of the group which are shown
in Figure~\ref{fig:radgrad}.  It is seen in the top panel of that plot
that the surface number density of galaxies declines smoothly with
radius out to $\sim 1.8^{\circ}$ then abruptly drops a factor $\sim 4$
to a roughly constant plateau.  In the lower panel it is seen that the
velocity dispersion within radial shells drops by roughly a factor 2
inside $1.8^{\circ}$. Declining velocity dispersion profiles are
typical of dynamically evolved clusters
\nocite{Mahdavi99,Biviano04,Sand04,Mahdavi04}({Mahdavi} {et~al.} 1999; {Biviano} \& {Katgert} 2004; {Sand} {et~al.} 2004; {Mahdavi} \& {Geller} 2004). Beyond 1.8\degr, locally
averaged velocity dispersions drop to the low levels see in the Local
Volume \nocite{Karachentsev03}({Karachentsev} {et~al.} 2003).

The properties demonstrated in Fig.~\ref{fig:radgrad} are as expected
from a collapsed group.  The $\sim 1.8^{\circ}$ dimension can be
inferred to mark the {\it caustic of second turn-around}
\nocite{Bertschinger85}({Bertschinger} 1985).  Objects bound to the group decouple
from the cosmic expansion at a radius of first turn-around (the
zero-velocity surface around the group) \nocite{Sandage86}({Sandage} 1986), collapse
then reexpand to a second turnaround, then continue to oscillate while
exchanging orbital energy with other group components.  An observable
cusp might be anticipated from the recent arrivals that have only had
time to pass once through the group and just reach second turnaround.
All other galaxies that have collapsed should lie interior (with rare
slingshot exceptions).

If one gives consideration to other dynamically evolved groups in the
Local Supercluster in a similar manner, one can infer a plausible
second turnaround caustic by looking for an outer boundary to the
distribution of early type systems.  One of the authors (RBT) has
undertaken this task, and has determined the following unpublished
relation between the group velocity dispersion, $\sigma_V$, and the
radius of the apparent caustic of second turnaround, $r_{2t}$, from
observations of 7 dense groups and clusters in the Local
Supercluster:
\begin{equation}
\sigma_V / r_{2t} = 390~{\rm km~s^{-1}~Mpc^{-1}}.
\end{equation}
The conditions in the NGC~5846 Group are consistent with this
relationship. Because each value of $r_{2t}$ can be used to define a
value of $\sigma_V(r_{2t})$ using the velocity dispersion of the
galaxies within $r_{2t}$, the above expression is a nonlinear equation
in one variable. Solving the equation numerically, we find that
\nvirial\ galaxies with velocities giving a dispersion of
$\sigma_V=320~{\rm km~s^{-1}}$, defining a circle of projected radius
$r_{2t}=0.84$~Mpc~$= 1.84^{\circ}$, are included within the projected
circle. This value of $r_{2t}$ is consistent with the radius at which
the velocity dispersion profile levels off (Figure \ref{fig:radgrad}).

The $r_{2t}$ circle is superimposed on Fig.~\ref{fig:presurvey}, with
the center (at 226.40,+1.79) chosen to optimize the $r_{2t}$
enclosure.  We conduct a virial analysis based on these \nvirial\
galaxies, with no luminosity weighting. We apply the median virial
mass estimator \nocite{Heisler85}({Heisler} {et~al.} 1985), which involves no luminosity
weighing and does not require the determination of a velocity or
spatial center. 10000 Monte Carlo simulations of the group are used to
derive the errors on the estimated mass. We find the virial mass to be
$8.3 \pm 0.3 \times 10^{13} M_{\odot}$.

To further explore the group dynamics, we consider all 100 galaxies
with known redshifts within 3\degr\ (1.8 Mpc) of NGC 5846, the galaxies
plotted in Fig.~\ref{fig:presurvey}.
We construct a map of the mean velocity and
velocity dispersion across the group. We define two functions on the
sky: $v_n(\alpha_{2000},\delta_{2000})$ is the mean velocity of the
$n$ closest group members to the position
$(\alpha_{2000},\delta_{2000})$; $\sigma_n(\alpha,\delta)$ is the
velocity dispersion of the $n$ closest members to that position. We
show $v_{10}(\alpha_{2000},\delta_{2000})$ and
$\sigma_{10}(\alpha_{2000},\delta_{2000})$ along with the RASS X-ray
contours in Figure \ref{fig:meansigmap}.

The maps of $v_{10}$ and $\sigma_{10}$ are complex in character.  
There is considerable structure in the velocity
dispersion across the face of the group. The regions associated with
the bright galaxies NGC 5846, NGC 5813, and NGC 5831 show large
velocity dispersions, $\approx 500$ km s\m, while the north-south
region at $\alpha_{2000} =$ 15:03:30 (exactly between the NGC 5846 and
NGC 5813 subgroups) shows a low velocity dispersion, $\approx 300$ km
s\m.  The spatial clumping and the velocity
dispersion signatures suggest that NGC~5846 and NGC~5813 are the centers 
of distinct sub-structures.   The persistence of this substructure is
interesting because the characteristic crossing-time between these
two centers (separation divided by mean group dispersion) is only
1.5~Gyr, $\sim 10\%$ the age of the universe.  

The velocity map $v_{10}$ shows a general north-south trend in the
mean local velocity of the galaxies: the galaxies to the north of the
NGC 5846--NGC 5813 axis have lower velocities than those to the
south, perhaps indicative of rotation.
The complexity of the dynamical maps could not have been predicted
from the featureless velocity histogram (Figure \ref{fig:velhist}b).
A single-peaked, regular velocity
distribution is not necessarily indicative of a relaxed system. 
The system has almost certainly not reached virial equilibrium.

\begin{figure}
\begin{center}
\resizebox{3.5in}{!}{\includegraphics{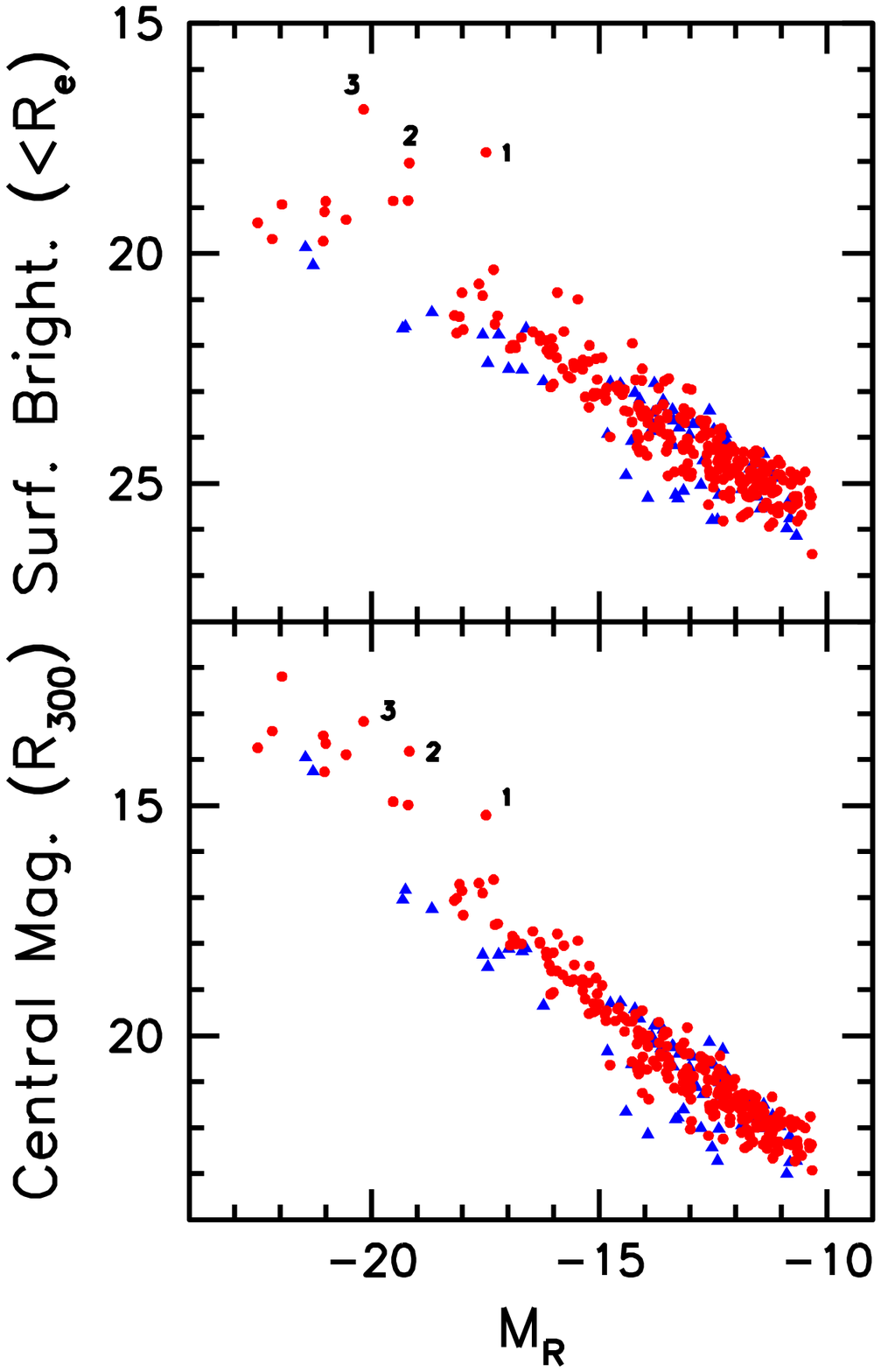}}
\end{center}
\figcaption{ \small 
{\it Top:} Mean surface brightness within the half-light radius as a
function of absolute magnitude.
Galaxies of early type are indicated by circles and galaxies of late
type are indicated by triangles.
{\it Bottom:} Central magnitude within an aperture of 300 pc 
($2.4^{\prime \prime}$)
radius as a function of absolute magnitude.  The numbers on the plots
identify [1] N5846--205, [2] NGC~5846A, and [3] NGC~5845.
\label{fig:sb-r300}}
\end{figure}

\begin{figure}
\begin{center}
\resizebox{3.5in}{!}{\includegraphics{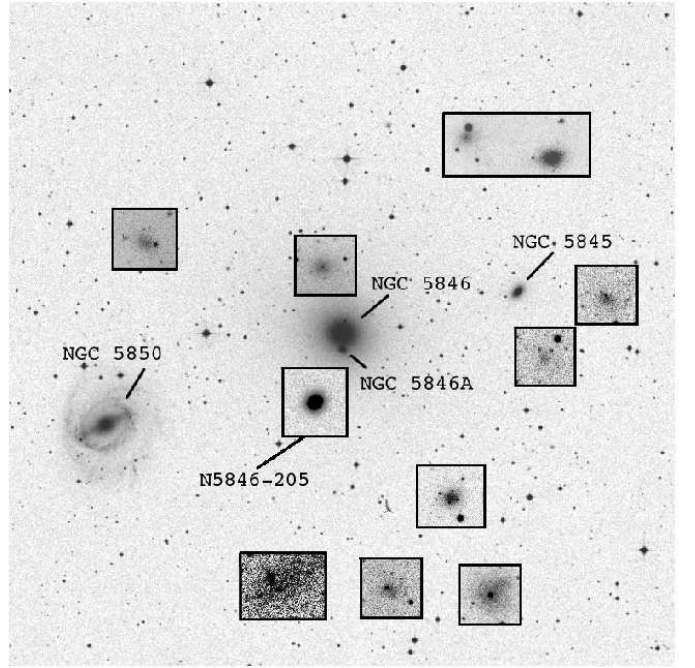}}
\end{center}
\figcaption{ \small The core of the group, with all likely members
(spectroscopically confirmed, priority 1, and priority 2
galaxies). The ultra compact dwarfs are N5846--205, NGC 5846A, and NGC
5845. The entire image is a $30\arcmin \times 30\arcmin$ (230 kpc
$\times$ 230 kpc) section of a POSS plate.  Each black box is a
zoomed-in view of a candidate dwarf galaxy as imaged by the CFHT. The
magnification within the black boxes is a factor of 5 compared to the
area outside the boxes.
\label{fig:coremap}}
\end{figure}

\section{Properties of the Member Galaxies}

\subsection{Surface Brightness Scaling Relations}

The surface brightness properties of the entire sample of confirmed
group members and plausible candidates is demonstrated in
Figure~\ref{fig:sb-r300}.  The top panel shows the dependence with
luminosity of the mean surface brightness within the radius containing
half the light of a galaxy, the effective radius, while the bottom
panel shows the dependence of the central magnitude through a 300 kpc
radius metric aperture.  In both plots there is a remarkably clean
separation between high and low surface brightness systems.  A modest
separation is also found with morphological type in these plots.
Types Sa and earlier are indicated by circles and lie slightly above
types Sab and later, labeled with triangles.

The two plots carry similar information about an overall decrease in
surface brightness in proceeding from giant galaxies to dwarfs.  There
is a small difference between the plots for systems toward the lower
luminosity end of the high surface brightness group.  If one splits
the high surface brightness sample at $M_R=-20.5$, it is seen that the
fainter portion have higher mean surface brightnesses within an
effective radius but marginally lower metric central magnitudes than
the brighter portion.  The increase in mean surface brightness is
consistent with the scaling relation found by \nocite{Kormendy77}{Kormendy} (1977).
\nocite{Trujillo01}{Trujillo} {et~al.} (2001) point out that the ratio of central to mean surface
brightness varies in a well correlated way with the \nocite{Sersic68}{S\'ersic} (1968)
parameterization of the radial distribution of light.  The increase in
mean surface brightness toward fainter luminosities reflects a trend
in S\'ersic parameter.  As for the apparent decrease in the metric
central luminosity toward fainter though high surface brightness
galaxies, partly this could be an artifact of resolution.  The
$R_{300}$ parameter, the magnitude within a radius of 300~pc = 
$2.4^{\prime \prime}$
was chosen to represent the central flux without constraining to a
radius that would be affected by seeing.  However, the galaxy N5846--205 
at $M_R=-17.5$ (labeled 1 in the figure) has an effective radius of only
215~pc = $1.6^{\prime \prime}$ so the $R_{300}$ measure under-represents 
the central
flux.  With NGC~5846A and NGC~5845 (labeled 2 and 3 respectively),
effective radii are $\sim 500$~pc.  These unusual objects are given
attention in the following section.

\subsection{Small High Surface Brightness Objects}

The galaxies that are identified by numbers in Fig.~\ref{fig:sb-r300}
are remarkable for their high central densities and small dimensions.
They are N5846--205 (1), NGC 5846A (2), and NGC 5845 (3).  They only
come to our attention through spectroscopic confirmation of an
appropriate redshift.  Their high surface brightnesses would exclude
them from our sample of group candidates.  Hence, our selection
criteria demonstrably fails in this occasion occasion.  Is this
failure common or rare?

One could ask if there is a relationship with the ultracompact dwarfs
that have been found in the Fornax Cluster \nocite{Phillips01}({Phillipps} {et~al.} 2001).  Those
objects are faint ($M_R \sim -12$) and small ($R_e \sim 20$~pc), so
small that they are indistinguishable from stars in ground-based
imaging.  They also only came to attention through spectroscopy.  In
the case of the NGC~5846 Group, SDSS spectroscopy has provided
reasonable completion of {\it non-stellar} targets in the field that 
are brighter than $R=17$ ($M_R=-15$).  Hence, an ultracompact
dwarf population like that found in Fornax would not be accessed with
the current observations in the NGC~5846 region, because such dwarfs 
would be both too faint and too small.

Still, the SDSS spectroscopy does extend to a faintness limit that is
interesting ($M_R \sim -15$) and precludes that there is a numerically
important population of high surface brightness objects in the range
$-15 > M_R > -19$.  Only one high surface brightness galaxy,
N5846--205, is found in this range.  This system has properties
similar to the Local Group elliptical M32, though N5846--205, with
$M_R=-17.5$, is brighter by a magnitude.  The other two high surface
brightness objects that draw attention in Fig.~\ref{fig:sb-r300}, NGC
5846A and NGC 5845, are small in size but have luminosities that put
them above the dwarf regime.

These three high surface brightness systems are very close to the core
of the group!  Their locations are seen in Figure~\ref{fig:coremap}.
These three galaxies are the closest spectroscopically confirmed group
members to NGC~5846, lying at 5, 25, and 55 kpc in projection (the
other galaxies shown in blowup boxes in Figure \ref{fig:coremap} are
low surface brightness candidate dwarfs without known
velocities). This close proximity to the dominant NGC~5846 of the
three most extreme objects in the surface brightness plot suggests
very strongly that these objects have been tidally stripped.  In this
respect, these objects may be very large counterparts to the Fornax
ultra compact dwarfs, suspected also to be the victims of tidal
stripping or `threshing' \nocite{Bekki03}({Bekki} {et~al.} 2003).

These objects are interesting in their own right but they are a distraction 
from the thread of the current investigation.  The SDSS velocity information
assures us that high surface brightness galaxies constitute only a tiny
fraction of the group population in the interval $-15 > M_R > -19$.
There is no information, but no reason to suspect, that high surface
brightness systems make up an important fraction of the population
fainter than $M_R=-15$.

\begin{figure*}
\begin{center}
\resizebox{7in}{!}{\includegraphics{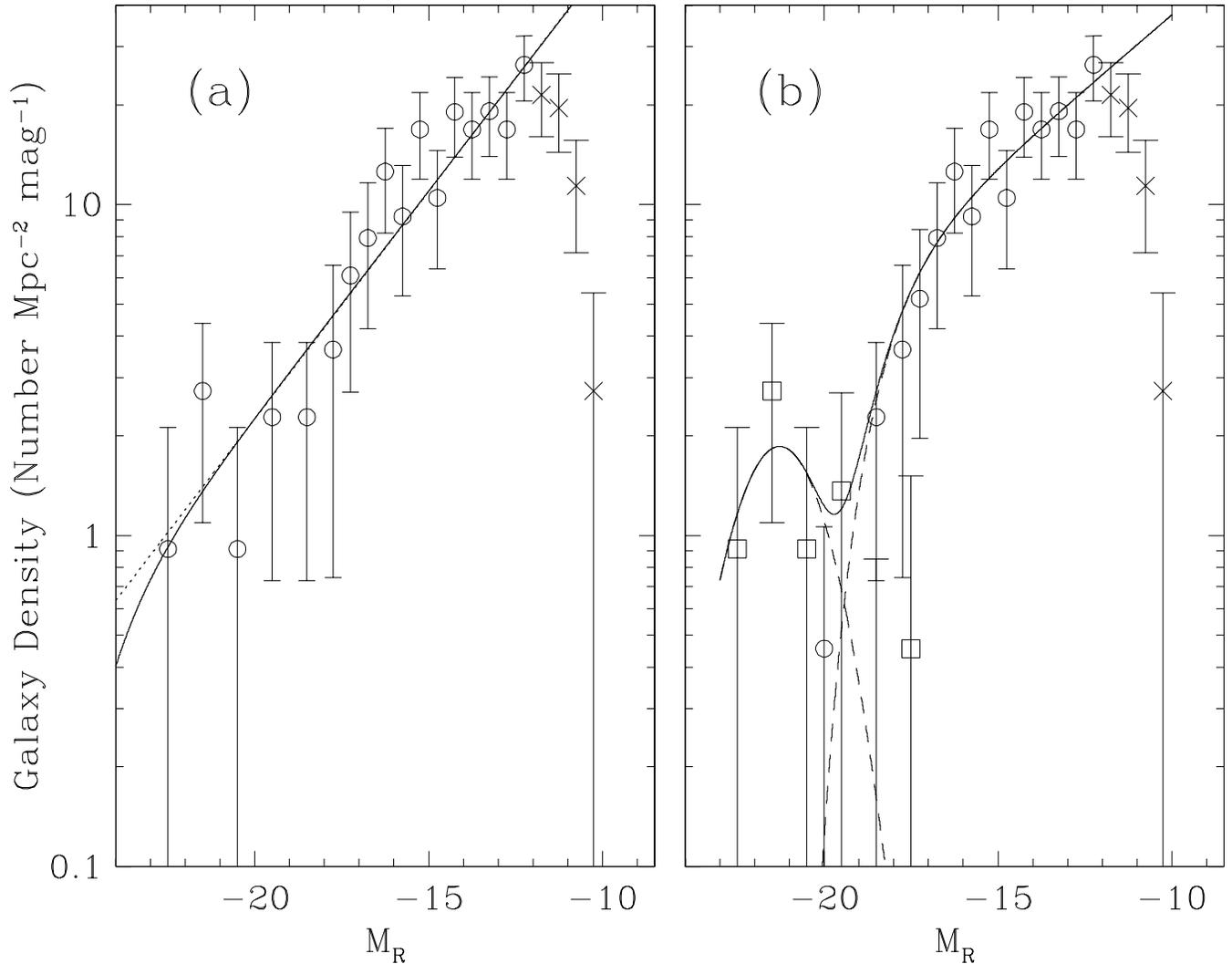}}
\end{center}
\figcaption{ \small  NGC~5846 Group luminosity function.  (\emph{a}) Circles
represent the density of likely group members. The solid line shows
the best-fit Schechter function, and the dotted line is the best-fit
single power law. (\emph{b}) Speculative reinterpretation of
(\emph{a}), with a Gaussian fit to the high surface brightness
galaxies (squares), and a Schechter function fit to the low surface
brightness galaxies (circles). The dotted lines show the individual
fits, and the solid line shows the sum of the two components.  The
fit results are given in Table \protect\ref{tbl:lf}
\label{fig:lf}}
\end{figure*}

\begin{figure}
\begin{center}
\resizebox{3.5in}{!}{\includegraphics{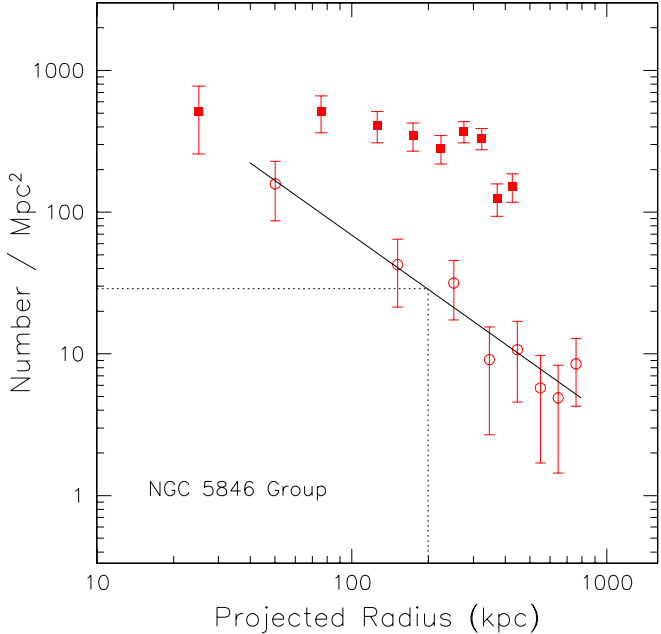}}
\end{center}
\figcaption{ \small 
The surface number density of galaxies in the NGC~5846 Group as a function of 
radius.  Filled symbols: all galaxies.  Open symbols: only galaxies with
$M_R < -17$.  The solid line is a fit to the radial distribution of bright 
galaxies.  The dotted lines locate the density at 200~kpc radius according
to this fit. 
\label{fig:density}}
\end{figure}

\subsection{Luminosity Function}

The analysis in the previous sections has provided strong constraints on
the group membership down to faint levels.   The group is sufficiently
populated that the domain of the region that has undergone collapse is
reasonably defined ($r_{2t}=0.84$~Mpc).  There is confirmation of
membership for essentially all galaxies brighter than $M_R=-15$ so the
luminosity function at the bright end is quite secure (though statistics
at high luminosities are limited).  At fainter magnitudes the partial 
velocity information and the spatial correlation information provide
good constraints on membership probabilities.  

To constrain the luminosity function of the galaxies, we place objects
in 0.5 mag bins and calculate the surface number density per unit
absolute magnitude. The result appears in Figure \ref{fig:lf}. If
there is only one galaxy in a bin, we combine that bin with an
adjacent one. The luminosity function is fit to a faint limit of
$M_R=-12$.  We assume that the dropoff faintward of that magnitude is
due to incompleteness. The final luminosity function includes galaxies
that are likely members according to the correlation analysis
(\S\ref{sec:membership2}): all galaxies confirmed by velocities as
members, plus all galaxies rated 0--2, plus 70\% of galaxies rated 3
brighter than $M_R=-13.4$, plus 50\% of galaxies of early type rated 3
fainter than $M_R=-13.4$. Because the number of galaxies per bin is
small, we use the \nocite{Gehrels86}{Gehrels} (1986) asymptotic formula for calculating
the error in a bin with $N$ members:
\begin{equation}
\epsilon_N = 1+(0.75+N)^{1/2}.
\label{eq:gehrels}
\end{equation}
This formula is accurate to 1.5\% for all values of $N$.

We find a striking feature: there is a plateau, or even dip, in the LF
at $M_R \sim -20$.  If real and not due to noise, this feature cannot
be described by a simple \nocite{Schechter76}{Schechter} (1976) function. A similar
feature exists, with greater significance, in the data described by
TT02. The Virgo cluster exhibits a quite significant and similar LF
``bump'' \nocite{TrenthamHodgkins02}({Trentham} \& {Hodgkin} 2002).

We attempt to model this luminosity function using progressively more complicated
fitting functions. To begin, in Figure \ref{fig:lf}a, we fit a simple
power law in luminosity,
\begin{equation}
N(M) = N_{-19} 10^{-0.4 (\alpha+1) (M + 19) },
\end{equation}
Where $N_{-19}$ is the number density of galaxies with $M_R = -19$,
and $\alpha$ is the slope defined so that it matches the faint-end
slope of a \nocite{Schechter76}{Schechter} (1976) function. We find this function to be an
acceptable fit to the data (see Table \ref{tbl:lf}). Given, however,
that a single power law mass or luminosity distribution is not
anticipated by either theoretical models of structure formation or 
observations, we also fit
a \nocite{Schechter76}{Schechter} (1976) function:
\begin{eqnarray}
\nonumber
N(M) = &  S(M) = & N_* \exp{\left[-10^{-0.4 (M - M_*)} \right]} \times \\
& & 10^{-0.4 (\alpha+1) (M - M_*) }
\end{eqnarray}
This function also yields an acceptable fit. However, the
characteristic magnitude $M_*$ is constrained to be $-24.0 \pm 3.2$ at
95\% confidence.  This value is only marginally consistent with other
analyses of the local galaxy luminosity function.  For example, the
overall SDSS luminosity function \nocite{Blanton01b}({Blanton} {et~al.} 2001) has $M_* =-20.8 \pm
0.3$ at the 68\% confidence level (the difference between the $R$ and
the $r$ band photometry is negligible given our error bars). The value
of $\alpha = -1.34 \pm 0.06$ is consistent with the overall SDSS
value. Figure \ref{fig:lf}a shows both the single power law and the
Schechter function fits to our data.

Because the single Schechter function fit yields somewhat too bright
an $M_*$, it is useful to evaluate other luminosity function
models. The surface brightness-magnitude diagram (Figure
\ref{fig:sb-r300}) suggests a third way to proceed. There appears to
be a significant gap in the surface brightness distribution in the
group population. An ensemble of high surface brightness galaxies are
clustered separately from the rest of the members (chiefly low-surface
brightness dwarfs).  We ask whether the two populations have separate
luminosity distributions, similarly to what is seen in TT02 and
\nocite{TrenthamHodgkins02}{Trentham} \& {Hodgkin} (2002). To test this idea, we fit these two
populations separately by dividing them into groups with $M_{R,300}<
16$ and $M_{R,300}> 16$. In the high surface brightness population,
one bin between $M_R = -19$ and $M_R = -17$ has zero members; the
error in this empty bin is still described by the \nocite{Gehrels86}{Gehrels} (1986)
formulation above (equation \ref{eq:gehrels}). The high surface
brightness population is fit with a Gaussian,
\begin{equation}
N(M) = N_g \exp{\left[ - \left(\frac{M - M_g}{\sigma_g} \right)
\right]^2},
\end{equation}
while the low-surface brightness population is fit with a Schechter
function. The results appear in Figure \ref{fig:lf}b and Table
\ref{tbl:lf}.

The Gaussian and Schechter functions provide a good fit to the LF of
the two respective populations. The faint-end slope $\alpha = -1.23 \pm
0.12$ is consistent with the faint-end slope of the single Schechter
function fit; the difference derives from the fact that the
high-luminosity galaxies are excluded, leading to a much fainter $M_*
= -18.5 \pm 1.3$. The well-known correlation between $M_*$ and
$\alpha$ then leads to a smaller best-fit value of $\alpha$. 

Our method for the inclusion of the priority 3 members may lead to
systematic errors in $\alpha$. To constrain these uncertainties, we
undertake two further fits. First, we measure $\alpha$ for only the
spectroscopically confirmed plus the priority 1-2 members, obtaining
$\alpha = -1.17 \pm 0.15$. Then we measure $\alpha$ for all the
spectroscopically confirmed members plus the priority 1-3 members,
obtaining $\alpha = -1.38 \pm 0.10$. Thus, we expect both the
statistical and systematic errors in $\alpha$ to equal be $\approx
0.1$. Our final adopted $\alpha$ is an attempt to encapsulate all the
complexities of the sample into a single value. We take the mean
between the whole-sample and the low surface brightness fits: $\alpha
= -1.3 \pm 0.1$ (statistical) $\pm 0.1$ (systematic).

Given the large number of free parameters and the small number of data
points, the need for a Gaussian high-luminosity component is not
statistically significant.  However, a formulation more complicated
than the Schechter function is required to describe any saddle between
the giants and the dwarfs, a saddle that it seen recurrently in
different samples, e.g. TT02 and \nocite{TrenthamHodgkins02}{Trentham} \& {Hodgkin} (2002).  The
present formulation may have some physical sense if the separation
between high and low surface brightness systems seen in the scaling
relations has some meaning.  We have discussed surface brightness
bimodality and a possible dynamical interpretation in the context of a
sample dominated by late-type disk galaxies \nocite{Tully97}({Tully} \& {Verheijen} 1997).  It is
not clear if that discussion has relevance to this sample of
predominantly early types.  We do not argue that the present sample in
itself provides justification for adopting a two-component luminosity
function.  The matter is something to review as material accumulates
for more environments.

\begin{figure}
\begin{center}
\resizebox{3.5in}{!}{\includegraphics{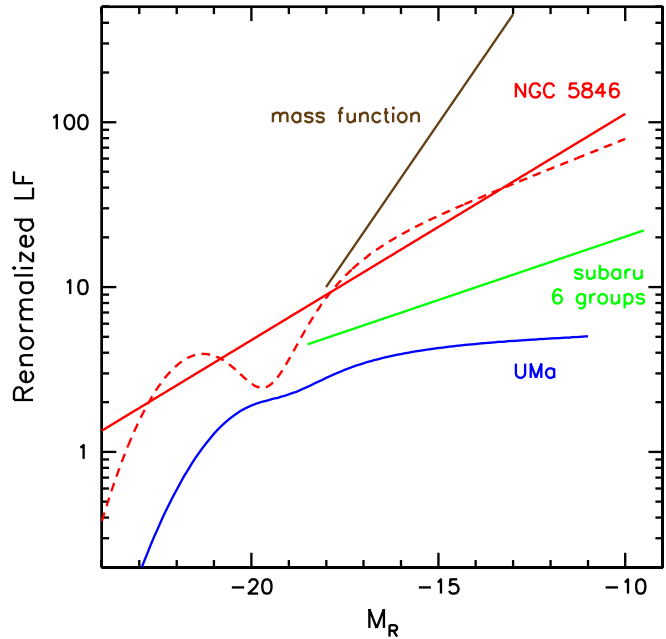}}
\end{center}
\figcaption{ \small Schematic comparison of the NGC~5846 Group
luminosity function with other samples.  A solid curve represents the
best-fit power law to the NGC~5846 Group. The dashed curve shows the
speculative Gaussian plus Schechter fit .  The curves labeled `subaru
6 groups' and `UMa' are taken from TT02.  The 6 groups curve
represents an average found for the dwarf regime in 6 groups discussed
by TT02.  The density normalizations are carried out in a consistent
way in these separate cases.  The steeper line labeled `mass function'
indicated the low mass slope of a modified Press-Schechter
distribution with arbitrary vertical scaling.
\label{fig:lf_compare}}
\end{figure}

Finally, we wish to compare the luminosity function for this group
with those found in other environments.  To make this comparison
we need a number density normalization.  As a matter of convenience
and to establish a convention, we determine a density in luminous
galaxies ($M_R < -17$) at a radius of 200~kpc from the group center.
We are looking for variations in the bright/faint distributions which
is why the group density definition is restricted to just the bright
members.  The metric radius of 200~kpc is chosen because it is
representative of the core of our smallest groups.  Details of the
normalization procedure are discussed by TT02.
Figure~\ref{fig:density} presents the run of the density of galaxies
with radius in the NGC~5846 Group in a fashion analogous to what is
found in TT02.  The fit to the density distribution of the luminous
galaxies (open symbols) gives a normalization of 29 galaxies/Mpc$^2$
at a group radius of 200~kpc. The resulting renormalized LF, and a
comparison with other groups, appears in Figure \ref{fig:lf_compare}.

\section{Summary}

\begin{figure}
\begin{center}
\resizebox{3.5in}{!}{\includegraphics{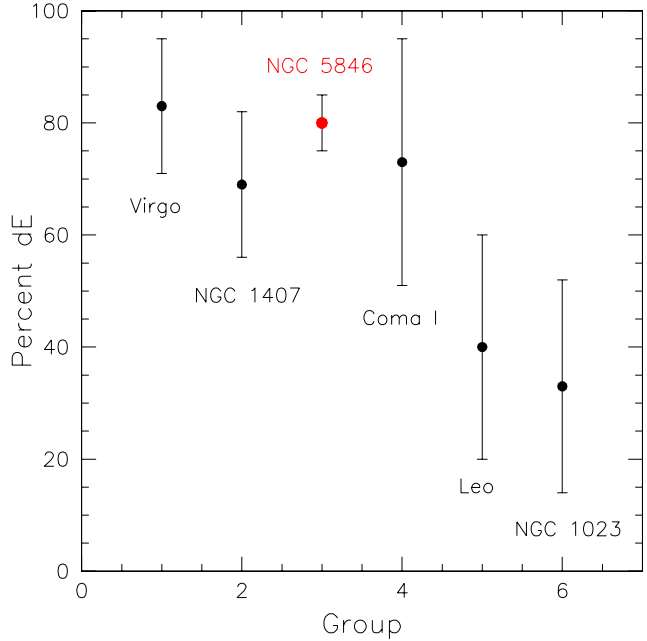}}
\end{center}
\figcaption{ \small Percentage of dwarf galaxies ($-17 < M_R < -11$)
classified as dE (including nucleated dE,N subtypes and transition
dE/I subtypes).  Data for groups other than NGC~5846 are from
TT02. The x-axis shows the ranking of the groups in order
of decreasing central galaxy density.
\label{fig:pcE}}
\end{figure}

The long-term goal of this project is to explore the nature of the
luminosity function of galaxies over a wide range of environments.  In
previous studies of environments in the Local Supercluster we have not
had the areal coverage to encompass substantial fractions of the
target groups.  This deficiency is corrected in the present study of
the NGC~5846 Group.  It is seen in Fig.~\ref{fig:presurvey} that the
area of our CFHT wide field survey covers almost the entire region
subsequently considered to lie within the second turnaround cusp,
$r_{2t}$.  Dwarf galaxies as faint as $M_R \sim -10$ can be identified
in the group, with incompletion setting in at $M_R \sim -12$.

Almost all galaxies associated with the group with $M_R < -15$ are
spectroscopically confirmed.  We identify \ntotal\ probable or
plausible candidates of which \nspecmemb\ have redshifts.  Based on
redshift sampling and a correlation analysis, we suggest that $251 \pm
10$ of the \ntotal\ candidates are true members.  The number of dwarfs
in the NGC~5846 Group is very large.  If, following TT02, a
dwarf-to-giant ratio is defined as No. galaxies with $-11 > M_R > -17$
over No. galaxies with $M_R < -17$ then \emph{dwarfs / giants} $= 7.3
\pm 0.7$.  The error includes the uncertainty in the rating 3
memberships and incompletion near the $M_R=-11$ limit.  This ratio of
dwarfs to giants is larger than the values seen in any of the TT02
groups.  Figure~\ref{fig:pcE} compares the percentage of dwarf
elliptical galaxies ($-11 > M_R > -17$) in the NGC 5846 Group with the
TT02 groups.  The NGC~5846 Group is overwhelmingly populated by early
type dwarfs, with $80\% \pm 5\%$, comparable to the situation in the
Virgo Cluster.  About $1/3$ of the early-type dwarfs are nucleated.
 
The large dwarf to giant ratio is reflected in the relatively steep
luminosity function at faint magnitudes.  The simplified single power
law slope of $\alpha_d = -1.34 \pm 0.08$ is significantly steeper than
the mean slope for 6 groups of $\alpha_d = -1.19 \pm 0.06$ found by
TT02 ($95\%$ probability).  We also show the luminosity function found
for the low density, spiral rich Ursa Major Cluster where the faint
end slope is $\alpha_d \sim -1.0$.  All of these observed luminosity
functions are much shallower than the modified Press-Schechter mass
function expected from the $\Lambda CDM$ hierarchical clustering
paradigm \nocite{Sheth99}({Sheth} \& {Tormen} 1999).

The present observations taken with the earlier work already strongly
support the proposition that the faint end of the luminosity function
of galaxies varies with environment.  Alternatively expressed, the
ratio of dwarf to giant galaxies varies with environment.  The dense,
dynamically evolved NGC~5846 Group has a high dwarf/giant ratio and a
relatively steep faint end luminosity function.  Still, even in this
environment there is a dearth of dwarfs compared to the expectations
of the $\Lambda CDM$ mass spectrum.  There is the implication that
astrophysical processes have affected the visible manifestations of
low mass halos, and in ways that are more effective at suppression of
light in lower density environments.

\bigskip
We thank the anonymous referee for insightful comments. This program
involves observations with the Canada-France-Hawaii and Keck
telescopes.  It is supported by NSF award AST-03-07706.

\clearpage

\bibliography{}




\input{./datatable}

\begin{deluxetable}{cccccc}
\tablecaption{Membership Determination \label{tbl:members}}
\tablewidth{0in}
\tablehead{\colhead{Priority} & \colhead{Early-Type} & \colhead{Number of} & \colhead{Confirmed Members}& \colhead{Confirmed Members} & \colhead{Confirmed}\\
           \colhead{Rating}   & \colhead{Fraction} & \colhead{Candidates}& \colhead{In the Literature} & \colhead{With Keck I} & \colhead{Non-members}}
\startdata
0    &    0.60 &   35  &   35 &   0  & 0 \\
1    &    0.79 &   39  &    3 &   4  & 0 \\
2    &    0.90 &   102 &    6 &  13  & 0 \\
3    &    0.73 &   149 &   15 &   2  & 7 \\
4,5  & \nodata &\nodata&    6 &   0  & \nodata \\
\enddata
\end{deluxetable}

\begin{deluxetable}{lrr}
\tablecaption{NGC 5846 and 5813: X-ray and Optical Fits
\label{tbl:fits}}
\tablewidth{0in}
\tablehead{\colhead{Parameter} & 
\colhead{NGC 5846} & \colhead{NGC 5813}}
\startdata
\mc{3}{c}{Optical Fit using $r^\gamma e^{-b (r/r_e)^\alpha}$} \\
$\chi^2 / \nu$          & 216/234           & 285/293 \\
$\gamma$                &    =0      & $-1.23 \pm 0.01$\\
$\alpha$                & $0.253\pm 0.003$  & $1.01 \pm 0.06$ \\
50\% light radius($r_e$)& $8.8\pm 0.7$ kpc  & $10.7 \pm 0.6$ kpc\\
90\% light radius\tn{a} & $48\pm 5$ kpc     & $22 \pm 1$ kpc \\
\\
\mc{3}{c}{X-ray Fit using $\beta$ Models} \\
$\chi^2/\nu$            &  120/119          &  8/17 \\
$r_1$                   & $1.7 \pm 0.1$ kpc & $ 20 \pm 3$ kpc\\
$\beta_1$               & $2.2 \pm 0.6$     & $0.67 \pm 0.03$ \\
$r_2$                   & $28 \pm 4$ kpc    & \nodata \\
$\beta_2$               & $0.55 \pm 0.01 $  & \nodata \\
$\Sigma_2/\Sigma_1$     & $9.5 \pm 0.8$     & \nodata \\
50\% X-ray radius\tn{a} & $11.8 \pm 2.3$ kpc  & $33 \pm 8$ kpc\\
90\% X-ray radius\tn{a} & $>110$ kpc    & $180 \pm 80$ kpc\\
\enddata
\tablecomments{Best-fit values and 68\% confidence intervals are 
shown.}
\tablenotetext{a}{These parameters depend entirely on the
other fit parameters; the quoted intervals are derived from 
the best fit parameters and errors using Monte Carlo simulations.}
\end{deluxetable}

\begin{deluxetable}{lllr}
\tablecaption{Luminosity Function Fits \label{tbl:lf}}
\tablewidth{0in}
\tablehead{
\colhead{Data Set} & \colhead{Fit Type} & \colhead{Parameter} & \colhead{Best Fit}}
\startdata
All Data & Power Law & $\alpha$ &  $-1.34 \pm 0.08$\\
& $\chi^2/\nu$ = 8.2/15 & $N_{-19}$ & $3.0 \pm 1.2$\\
\\
All Data & Schechter & $\alpha$ &  $-1.34 \pm 0.08$\\
& $\chi^2/\nu=8.2/15$ & $M_{*}$  &  $-24.0 \pm 3.1$\\
& & $N_{*}$ & $0.49^{+5}_{-0.23}$\\
\\
Low-SB & Schechter & $\alpha$ & $-1.22 \pm 0.12$ \\
Galaxies & $\chi^2/\nu$ = 4.4/11 & $M_{*}$ & $-18.5 \pm 1.3$ \\
& & $N_{*}$ & $6.0 \pm 3.1$ \\
\\
High-SB & Gaussian & $M_G$ & $-21.2 \pm 1.6$ \\
Galaxies & $\chi^2/\nu = 1.1/4$& $\sigma_G$ & $1.8^{+5}_{-1.8}$ \\
& & $N_G$ & $1.9^{+13}_{-1.9}$ \\ 
\enddata
\tablecomments{The 95\% confidence intervals take the correlation
among all the parameters into account; they are reported as one-sided
when the lower and upper errors are similar, and as two-sided when
the lower and upper errors differ substantially. The normalizations $N_*$
and $N_{-19}$ are in units of Galaxies Mpc$^{-2}$ mag$^{-1}$.}
\end{deluxetable}

\end{document}

%% file: defs.tex
\newcommand{\mr}[1]{\mathrm{#1}}
\newcommand{\ASCA}{\emph{ASCA}}
\newcommand{\ROSAT}{\emph{ROSAT}}
\newcommand{\as}{^{\prime\prime}}
\newcommand{\betamodel}{$\bfit$-model}
\newcommand{\bfit}{\beta}
\newcommand{\btrue}{\beta_\mr{true}}
\newcommand{\chandra}{\emph{Chandra}}
\newcommand{\clusterrange}{0-0.9}
\newcommand{\cross}{\mb{\times}}
\newcommand{\dd}[2]{\frac{d {#1}}{d {#2}}}
\newcommand{\dlndln}[2]{\frac{d \, {\ln{#1}}}{d \, {\ln{#2}}}}
\newcommand{\tdlndln}[2]{d \, \ln{#1} / \, d \, \ln{#2} }
\newcommand{\dln}[2]{\frac{d \, {\ln{#1}}}{d {#2}}}
\newcommand{\dndn}[3]{\frac{d^{#3} {#1}}{d {#2}^{#3}}}
\newcommand{\ee}[1]{$\times 10^{#1}$}
\newcommand{\ene}{\mathcal{E}}
\newcommand{\einstein}{\emph{Einstein}}
\newcommand{\etal}{{et al.}\ }
\newcommand{\fism}{f_\mr{ISM}}
\newcommand{\grad}{\mb{\nabla}}
\newcommand{\grouprange}{1.5-1.9}
\newcommand{\hhh}{h_{100}}
\newcommand{\hsq}{h_{100}^{-2}}
\newcommand{\lb}{L_\mr{B}}
\newcommand{\lism}{L_\mr{ISM}}
\newcommand{\lsoft}{L_\mr{soft}}
\newcommand{\lts}{\!\!}
\newcommand{\lx}{L_\mr{X}}
\newcommand{\mb}[1]{\mathbf{#1}}
\newcommand{\mc}{\multicolumn}
\newcommand{\tn}{\tablenotemark}
\newcommand{\mdot}{\dot{M}}
\newcommand{\m}{$^{-1}$}
\newcommand{\ngalbin}{30}
\newcommand{\ngoodgal}{1020}
\newcommand{\ngoodgr}{33}
\newcommand{\ngood}{72}
\newcommand{\ngroup}{$41$}
\newcommand{\nqa}{50}
\newcommand{\nqb}{22}
\newcommand{\nqc}{188}
\newcommand{\ntotspec}{2424}
\newcommand{\ntot}{260}
\newcommand{\nunique}{28}
\newcommand{\p}{^\prime}
\newcommand{\rhodm}{\rho_\mr{dm}}
\newcommand{\rosat}{\emph{ROSAT}}
\newcommand{\vbar}{\langle v \rangle}
\newcommand{\xmm}{\emph{XMM}}
\newcommand{\normalmargins}{\oddsidemargin 0in \evensidemargin 0in
\topmargin 0.5in \textwidth 6.5in \textheight 9in}

%% file: datatable.tex
\begin{deluxetable}{crrrrrrrrrrrl}
\tabletypesize{\small}
\tablecaption{Photometric and Spectroscopic Data \label{tbl:data}}
\tablewidth{0in}
\tablehead{\colhead{Galaxy ID} & \colhead{P} & \colhead{$\alpha_{2000}$} & \colhead{$\delta_{2000}$} & \colhead{Type} & \colhead{$m_{R,25.2}$} & \colhead{$m_{R,300}$} & \colhead{$M_R$}  & \colhead{$A_R$} &  \colhead{$r_e$} & \colhead{$ \mu_R $} & \colhead{$c z$} & \colhead{Notes}  \\
\colhead{N5846-}\\
}
\startdata
001 & 3 & 14:57:32.5 & +00:49:40 & dE,N\ & 21.08 & 22.54 & -11.13 &  0.12 &  3.3 & 25.65 & \nodata & \nodata       \\
002 & 3 & 14:57:40.3 & +02:46:40 & dE\ & 20.77 & 21.74 & -11.45 &  0.13 &  2.6 & 24.85 & \nodata & \nodata         \\
003 & 3 & 14:57:43.3 & +01:21:18 & dI\ & 18.83 & 20.67 & -13.37 &  0.12 &  4.9 & 24.27 & \nodata & \nodata         \\
004 & 2 & 14:57:45.1 & +03:25:56 & dE\ & 20.37 & 21.83 & -11.85 &  0.14 &  3.7 & 25.22 & \nodata & \nodata 	      \\
005 & 3 & 14:57:48.5 & +02:11:47 & dE\ & 21.16 & 21.92 & -11.06 &  0.14 &  1.9 & 24.60 & \nodata & \nodata 	      \\
006 & 3 & 14:57:52.2 & +03:03:56 & dI\ & 18.82 & 20.22 & -13.38 &  0.12 &  3.5 & 23.50 & \nodata & \nodata 	      \\
007 & 3 & 14:57:52.3 & +02:26:39 & dE\ & 21.52 & 22.73 & -10.70 &  0.14 &  2.6 & 25.57 & \nodata & \nodata 	      \\
008 & 3 & 14:57:53.0 & +01:04:48 & dE\ & 20.14 & 21.11 & -12.07 &  0.12 &  2.8 & 24.37 & \nodata & \nodata 	      \\
009 & 0 & 14:57:53.1 & +00:56:03 & E/S0 & 15.35 & 17.90 & -16.86 &  0.12 &  8.9 & 22.09 & 1886 & \nodata 	      \\
010 & 3 & 14:58:07.0 & +02:03:43 & dE & 18.70 & 20.23 & -13.51 &  0.13 &  4.0 & 23.68 & \nodata & \nodata 	      \\
011 & 1 & 14:58:12.6 & +01:59:39 & dE,N\ & 19.81 & 21.27 & -12.39 &  0.12 &  3.6 & 24.62 & \nodata & \nodata       \\
\enddata
\tablecomments{The full version of the table will appear in the
Astronomical Journal. The morphological type is indicated as follows:
dI, dwarf irregular; dE, dwarf elliptical; N, nucleated dwarf; E,
elliptical; S, spiral; and VLSB, very low surface brightness galaxy. P
is the original membership probability assigned to the galaxy purely
from photometry and morphology (0 = non-SDSS spectroscopic redshift; 1
= probable member; 2 = possible member; 3 = conceivable member; 4 =
likely not a member). $m_{R,25.2}$, $m_{R,300}$ and $M_R$ are the
measured $R$ band magnitudes, with the first being the isophotal
magnitude (to 25.2 mag arcsec\m), the second being the magnitude
within 300 pc (2.3\arcsec), and the third being the isophotal absolute
magnitude. $A_R$ is the reddening, and $r_e$ is the half-light radius
in arcseconds.  The heliocentric velocity in km s\m\ is $c z$.  The
notes give other galaxy identifications; ZM indicates data from
\protect\cite{Zabludoff98}.  All redshifts are from the NASA
Extragalactic Database, unless the date of a Keck observation is
indicated, in which case they are new. Galaxies marked with ``sat''
were saturated in our imaging; for these galaxies, all values were
derived from SDSS public data, with appropriate conversions from the
SDSS $r$ to Cousins $R$ photometric bands.}
\end{deluxetable}